\pdfoutput=1

\documentclass[a4paper, 12pt]{article}
\usepackage[a4paper]{geometry}
\usepackage{amsmath,amssymb,amsthm}
\usepackage[ruled, boxed, linesnumbered]{algorithm2e}
\usepackage{xcolor, colortbl}
\usepackage{caption}
\usepackage{subcaption}
\usepackage{latexsym}
\usepackage{graphics}
\usepackage{graphicx}
\usepackage{wasysym}
\usepackage{stmaryrd}
\usepackage[english]{babel}
\usepackage{url}
\usepackage{relsize}
\usepackage{authblk}

\usepackage{tikz}
\usetikzlibrary{shapes,arrows,backgrounds,patterns}
\tikzstyle{place}=[circle,draw=blue!50,fill=blue!20,thick]
\tikzstyle{placer}=[circle,draw=blue!50,fill=red!40,thick]
\tikzstyle{placeg}=[circle,draw=blue!50,fill=green!40,thick]
\tikzstyle{placey}=[circle,draw=blue!50,fill=yellow!60,thick]

\allowdisplaybreaks
\author[1,2]{Martin Ritchie}
\author[2]{Luc Berthouze}
\author[1]{Istvan Z. Kiss \thanks{Corresponding author: I.Z.Kiss@sussex.ac.uk}}
\affil[1]{School of Mathematical and Physical Sciences, Department of Mathematics, University of Sussex, Falmer, Brighton BN1 9QH, UK}
\affil[2]{Centre for Computational Neuroscience and Robotics, University of Sussex, Falmer,  Brighton, BN1 9QH, UK}

\def\HiLiy{\leavevmode\rlap{\hbox to \hsize{\color{yellow!30}\leaders\hrule height .8\baselineskip depth .5ex\hfill}}}
\def\HiLig{\leavevmode\rlap{\hbox to \hsize{\color{green!20}\leaders\hrule height .8\baselineskip depth .5ex\hfill}}}
\def\HiLio{\leavevmode\rlap{\hbox to \hsize{\color{orange!50}\leaders\hrule height .8\baselineskip depth .5ex\hfill}}}
\def\HiLib{\leavevmode\rlap{\hbox to \hsize{\color{blue!20}\leaders\hrule height .8\baselineskip depth .5ex\hfill}}}
\def\HiLir{\leavevmode\rlap{\hbox to \hsize{\color{red!20}\leaders\hrule height .8\baselineskip depth .5ex\hfill}}}
\definecolor{LightBlue}{rgb}{0.88,0.95,1}
\begin{document}

%----------------------------------------------------------------------------------------
\title{\bf Generation and analysis of networks with a prescribed degree sequence and subgraph family: Higher-order structure matters \\}
%----------------------------------------------------------------------------------------

\maketitle

\begin{abstract}
Designing algorithms that generate networks with a given degree sequence while varying both subgraph composition and distribution of subgraphs around nodes is an important but challenging research problem. Current algorithms lack control of key network parameters, the ability to specify to what subgraphs a node belongs to, come at a considerable complexity cost or, critically, sample from a limited ensemble of networks. To enable controlled investigations of the impact and role of subgraphs, especially for epidemics, neuronal activity or complex contagion, it is essential that the generation process be versatile and the generated networks as diverse as possible. In this paper, we present two new network generation algorithms that use subgraphs as building blocks to construct networks preserving a given degree sequence. Additionally, these algorithms provide control over clustering both at node and global level. In both cases, we show that, despite being constrained by a degree sequence and global clustering, generated networks have markedly different topologies as evidenced by both subgraph prevalence and distribution around nodes, and large-scale network structure metrics such as path length and betweenness measures.  Simulations of standard epidemic and complex contagion models on those networks reveal that degree distribution and global clustering do not always accurately predict the outcome of dynamical processes taking place on them. We conclude by discussing the benefits and limitations of both methods. 
\end{abstract}

\newpage

%%%----------------------------------------------------------------------------------------
%	INTRODUCTION
%----------------------------------------------------------------------------------------

\section{Introduction}\label{sec:intro}

Being able to replicate, and therefore investigate, the structure and function of real-world complex networks is a profoundly difficult problem. However, the pervasiveness of systems that could be more accurately interpreted as a result cannot be overstated: social networks  \cite{girvan2002community}, the spread of disease \cite{newman2003structure}, artificial intelligence \cite{hopfield1982neural}, language structure \cite{ronen2014links}, and transportation networks \cite{santi2014quantifying}. Accordingly, a number of network models and network generating algorithms have been proposed~\cite{watts1998collective, newman2001random, kim2004performance, volz2004random, serrano2005tuning, bansal2009exploring, ritchie2014higher, ritchie2014beyond, overbury2015using}. Many of these network models seek to reproduce a specific network property or characteristic: the degree distribution \cite{barabasi1999emergence, newman2001random}, the small worldness \cite{watts1998collective}, degree-degree correlations \cite{newman2002assortative, newman2003structure} or clustering, the propensity of 3-cycles in a network \cite{milo2002network, newman2003properties}. However, investigations of \emph {higher-order} structure, subgraphs and arrangements of subgraphs not specified by standard network metrics, have been limited by a lack of accurate and versatile network models. Some progress has been made using the \emph{configuration model}~\cite{ karrer2010random,ritchie2014higher, ritchie2014beyond}, and it is this work we seek to build upon.  

\begin{figure}[!htbp]
\begin{center}
\begin{tabular}{ccccc}
\begin{tikzpicture}
\node [place] (2) at (0, 0) {};
\node [place] (1) at (0.001, 1) {};
 \draw [ultra thick](1) to (2);
\end{tikzpicture}
&
\begin{tikzpicture}
  \newdimen\R
  \R=0.58cm
  \draw (0:\R)
     \foreach \x in {120,240,360} { (\x:\R) }
               (360:\R) node [place] (3) {}
               (240:\R) node [place] (2) {}
               (120:\R) node [place] (1) {};
  %   \draw [ultra thick](1) to (2);
     \draw [ultra thick](2) to (3);
     \draw [ultra thick](3) to (1);
\end{tikzpicture} 
&
\begin{tikzpicture}
  \newdimen\R
  \R=0.58cm
  \draw (0:\R)
     \foreach \x in {120,240,360} { (\x:\R) }
               (360:\R) node [place] (3) {}
               (240:\R) node [place] (2) {}
               (120:\R) node [place] (1) {};
   %  \node [place] (4) at (0, 0) {};
     \draw [ultra thick](1) to (2);
     \draw [ultra thick](2) to (3);
     \draw [ultra thick](3) to (1);
\end{tikzpicture} 
&
\begin{tikzpicture}
\node [place] (1) at (0.001, 0) {};
\node [place] (2) at (0.001, 1.001) {};
\node [place] (3) at (1, 1) {};
\node [place] (4) at (1, 0.001) {};
%\draw [ultra thick](1) to (2);
\draw [ultra thick](2) to (3);
\draw [ultra thick](3) to (4);
\draw [ultra thick](4) to (1);
\end{tikzpicture}
&
\begin{tikzpicture}
  \newdimen\R
  \R=0.65cm
  \draw (0:\R)
     \foreach \x in {90,210,330} { (\x:\R) }
               (330:\R) node [place] (3) {}
               (210:\R) node [place] (2) {}
               (90:\R) node [place] (1) {};
     \node [place] (4) at (0, 0) {};
     \draw [ultra thick](1) to (4);
     \draw [ultra thick](2) to (4);
     \draw [ultra thick](3) to (4);
\end{tikzpicture} 
\\
$G_0$ &    
$u3$ &
$G_\triangle~/~t3$ & % $G_\boxempty~/~e4$
$u4$ & %$G_\boxslash~/~d4$
$s4$ \\  %$G_\boxtimes~/~c4$
\begin{tikzpicture}
\node [place] (1) at (0.001, 0) {};
\node [place] (2) at (0.001, 1.001) {};
\node [place] (3) at (1, 1) {};
\node [place] (4) at (1, 0.001) {};
\draw [ultra thick](1) to (2);
\draw [ultra thick](2) to (3);
%\draw [ultra thick](3) to (4);
\draw [ultra thick](4) to (1);
\draw [ultra thick](1) to (3);
\end{tikzpicture}
&
\begin{tikzpicture}
\node [place] (1) at (0.001, 0) {};
\node [place] (2) at (0.001, 1.001) {};
\node [place] (3) at (1, 1) {};
\node [place] (4) at (1, 0.001) {};
\draw [ultra thick](1) to (2);
\draw [ultra thick](2) to (3);
\draw [ultra thick](3) to (4);
\draw [ultra thick](4) to (1);
\end{tikzpicture}
&
\begin{tikzpicture}
\node [place] (1) at (0.001, 0) {};
\node [place] (2) at (0.001, 1.001) {};
\node [place] (3) at (1, 1) {};
\node [place] (4) at (1, 0.001) {};
\draw [ultra thick](1) to (2);
\draw [ultra thick](2) to (3);
\draw [ultra thick](3) to (4);
\draw [ultra thick](4) to (1);
\draw [ultra thick](1) to (3);
%\draw [ultra thick](2) to (4);
\end{tikzpicture} 
&
\begin{tikzpicture}
\node [place] (1) at (0.001, 0) {};
\node [place] (2) at (0.001, 1.001) {};
\node [place] (3) at (1, 1) {};
\node [place] (4) at (1, 0.001) {};
\draw [ultra thick](1) to (2);
\draw [ultra thick](2) to (3);
\draw [ultra thick](3) to (4);
\draw [ultra thick](4) to (1);
\draw [ultra thick](1) to (3);
\draw [ultra thick](2) to (4);
\end{tikzpicture} 
&
~
\\
$i4$ &    
$G_\boxempty~/~e4$ &
$G_\boxslash~/~d4$ &
$G_\boxtimes~/~c4$ &
			 \\ 
\begin{tikzpicture}
  \newdimen\R
  \R=0.75cm
  \draw (0:\R)
     \foreach \x in {1,71,...,281} { (\x:\R) }
               (1:\R)   node [place] (1) {}
               (71:\R)  node [place] (2) {}
               (141:\R) node [place] (3) {}
               (211:\R) node [place] (4) {}
               (281:\R) node [place] (5) {};
     \draw [ultra thick](1) to (2);
     \draw [ultra thick](2) to (3);
     \draw [ultra thick](3) to (4);
     \draw [ultra thick](4) to (5);
     \draw [ultra thick](5) to (1);
\end{tikzpicture}
&
\begin{tikzpicture}
  \newdimen\R
  \R=0.75cm
  \draw (0:\R)
     \foreach \x in {1,61,...,301} { (\x:\R) }
               (1:\R)   node [place] (1) {}
               (61:\R)  node [place] (2) {}
               (121:\R) node [place] (3) {}
               (181:\R) node [place] (4) {}
               (241:\R) node [place] (5) {}
               (301:\R) node [place] (6) {};
     \draw [ultra thick](1) to (2);
     \draw [ultra thick](2) to (3);
     \draw [ultra thick](3) to (4);
     \draw [ultra thick](4) to (5);
     \draw [ultra thick](5) to (6);
     \draw [ultra thick](6) to (1);
\end{tikzpicture}
&
~
&
~
&
\\
$G_{\pentagon}$&    
$G_{\hexagon}$&
~&
~& 
\end{tabular}
\caption{The set of subgraphs that have been used in this paper. The subgraphs denoted by: $\{G_0, G_\triangle, G_\boxempty, G_\boxslash, G_\boxtimes, G_{\pentagon}, G_{\hexagon}\}$, are those that have been used as input for the proposed network construction algorithms. We use: $\{u3, t3, u4, s4, i4, e4, d4, c4\}$, to denote the total number of uniquely counted subgraphs given by the subgraph counting algorithm~\cite{ritchie2014higher}.}\label{fig:examples}
\end{center}
\end{figure}
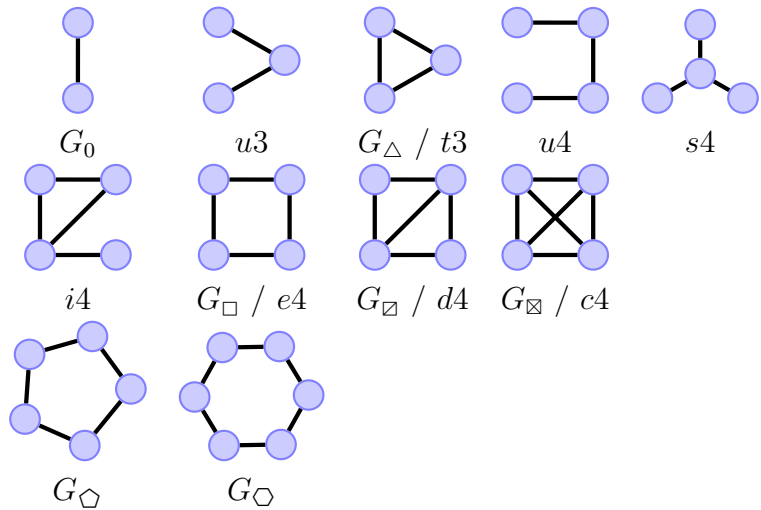

In the standard configuration model, triangle subgraphs appear infrequently as a by-product of working with finite size networks~\cite{bollobas1980probabilistic}. But what if one \emph{wants} triangle subgraphs to appear in a network, in particular, if one wants to model a complex network with clustering? An extension of the configuration model to this case exists~\cite{miller2009percolation, newman2009random}. In this extension a node is allocated a number of stubs, that may go on to form standard edges, as well as a number of triangle `corners' or \emph{hyperstubs}, pairs of stubs that will form triangles. While edges are formed in the usual way, triangles are formed by selecting three triangle hyperstubs at random and connecting their pairs of constituent stubs. As for edges the number of all stubs must be divisible by two, the total number of triangle hyperstubs must be divisible by three for the triangle hyperstub sequence to be graphical. Another similarity this model shares with the standard configuration model is that the probability that any two triangles will share an edge, thus forming a $G_\boxslash$ subgraph (see Figure~\ref{fig:examples}), vanishes in the limit of large network size~\cite{karrer2010random}. Just as a network composed of lines only is limited in recreating real-world networks, so is a model that can only include edges and triangles. Obviously, this may depend on properties and structure of the real networks, but in many cases edges and triangles are not enough to produce an accurate enough artificial replica of the real network. 

The configuration model has since received further attention to address this~\cite{karrer2010random}. Building on the edge-triangle model a more general subgraph-based approach is taken where one may specify distributions of edges alongside distributions of arbitrary subgraphs. In the case of complete subgraphs it is obvious how to do this. For example, $G_\boxtimes$ subgraphs can be formed by allocating to nodes hyperstubs composed of three stubs. Then, four of these hyperstubs can be selected at random to form a $G_\boxtimes$ subgraph. However, it is not clear how this may work for subgraphs that are not symmetric. For example, in a $G_\boxslash$, there are two different types of hyperstubs and it is necessary for any network model or construction algorithm to be able to make this distinction. Karrer and Newman proposed that it is possible to identify a node's \emph{role} within a subgraph using \emph{orbits}. To find the orbits of a subgraph one must first list all possible automorphisms of the subgraph, that is, permutations of nodes that do not create or destroy edges. The orbit of a node is a set of other nodes with which it may be permuted so that no edges are created or destroyed. Of course, computing the automorphism group of subgraphs is a computationally hard problem but so long as subgraphs with few nodes are used, this is not a problem~\cite{karrer2010random}. 

Network models are rarely used independently of other processes. Instead, they typically provide the substrate for dynamical processes to operate upon. For example, the compartmental Susceptible-Infected-Recovered ($SIR$) model of contagion is often embedded into a network to help better understand how the network and its properties affect the epidemic. Previous work~\cite{ritchie2014beyond} successfully incorporated the Karrer and Newman approach into an approximate ODE or mean-field model for $SIR$ epidemics on networks displaying higher-order structure, and this mean-field model showed excellent agreement with simulation results. In order to achieve this, Ritchie et~al. bypassed the need to classify a node's role in a subgraph via the automorphism group. Instead, nodes within asymmetric subgraphs were uniquely enumerated, even if they were topologically equivalent to one another, and this enumeration defined their role. The motivation for this adaptation was to simplify the derivation of the ODE model. Using the orbit approach or the full enumeration are different ways of satisfying different modelling needs, and these are not the only possible approaches. In fact, when modelling networks and nodes within subgraphs one can instead classify nodes by the stub cardinality of their hyperstubs. 

A common method across all of the above models, i.e., edge-triangle, the more general Karrer-Newman model, and that proposed by Ritchie et al., is that sequences of hyperstubs must be specified for each and every subgraph that is to be included. From these sequences it is possible to recover the network's degree sequence by multiplying them by the stub cardinality of the hyperstub which they represent, and then summing the resulting sequences. Therefore the degree sequence of the network is a result of the construction of the network rather than a quantity that is controlled for. However, given that the degree sequence of the network is probably the single most important characteristic of a network, there is a need for methods that can generate networks with a particular subgraph family and distribution yet preserve a given degree sequence. In~\cite{ritchie2014beyond}, we recently showed that it is possible to constrain the  hyperstub sequences so that the $1^{st}$ and $2^{nd}$ moments of the resulting degree sequence are controlled. In this paper, we go beyond this work and propose two generation algorithms that provide full control over the degree sequence and clustering. 
 
The paper is organised as follows. In Section~\ref{sec:MM}, we describe in detail the two generation algorithms, including tuning of clustering. In Section~\ref{sec:results}, we validate our algorithms and we explore the diversity of the generated networks by comparing them to the widely used Big-V rewiring scheme. We further analyse networks generated by using different subgraph families or distributions. Epidemic and complex contagion models are simulated on these networks and we show that degree distribution and global clustering alone are not sufficient to predict the outcome of these processes. Finally, we discuss extensions and further research questions relating to our work.

\section{Materials and methods}\label{sec:MM}

In this section we propose two new algorithms, both of which are parametrised by a degree sequence and a set of subgraphs. The algorithms  construct hyperstub degree sequences (from which the input degree sequence may be recovered exactly) that can be used in a modified configuration model style connection procedure to realise a network. 

There are some caveats regarding the preservation of the input degree sequence that are common to all configuration-like models. Firstly a degree sequence must sum to a an even number to be graphical. If it does not, a stub must be created or destroyed to satisfy this constraint. In general, hyperstub degree sequences must sum with multiplicity equal to the number of times they appear in their parent subgraphs, i.e., a triangle hyperstub sequence must be divisible by 3. When selecting stubs or hyperstubs at random to form subgraphs it is possible that self or multi-edges may form. The number of these events happening depends only on the average degree $\langle k \rangle$ and thus remains constant with network size. It is possible to simply delete self-edges or collapse multi-edges down to a single edge. If this approach is taken then the guiding degree sequence will be violated. Instead we disallow such connections by reselecting nodes in the connection procedure until no self or multi-edges will be created by forming the subgraph. This is known as the \emph{matching algorithm}~\cite{milo2003uniform}. Finally, it is possible for the process to be left with no option other than to add subgraphs over existing links or selecting multiple instances of the same node. In this case we completely reset the algorithm, regenerating hyperstub sequences and forming subsequent connections until a network is formed. 

\subsection{The underdetermined sampling algorithm -- UDA}\label{sec:UDA}

The concept underpinning this algorithm is that for each node there are combinations of hyperstubs that will satisfy its degree. For example, a node with $k=3$ classical edges could form 3 single $G_0$ edges or 1 $G_0$ edge and one $G_\triangle$ hyperstub. The number of possible arrangements will depend on the degree of the node and number of input subgraphs. From these arrangements a single one is selected at random. For a given degree $k$ this problem is equivalent to solving an underdetermined linear Diophantine equation equal to $k$ subject to positivity constraints. The coefficients are given by the edge counts of the hyperstubs, that are induced by the input subgraphs, and the solution will give the number of each hyperstub so that the degree of the node is matched exactly.  

To generate a network using this algorithm, let us assume that a degree sequence, $D=\{d_1,d_2, \dots, d_{N}\}\in \mathbb{N}_0^{1 \times N}$, and the set of subgraphs to be included in the network's construction, $G = \{G_1, G_2, \dots, G_l\}$, is given. Then, for each subgraph we classify its hyperstubs by their edge cardinality. It is now possible to form a vector that has elements specifying the number of edges in each hyperstub. From this vector we take the unique elements. For example, the $G_\boxslash$ subgraph will have a corresponding hyperstub vector of $\alpha = (2,3)$. For a given degree $k$ we must consider all possible hyperstubs and hyperstub combinations that yield a classical degree equal to $k$. To systematically list all such combinations, we first concatenate all the hyperstub vectors into a single vector, $\boldsymbol{\alpha}$, to be used as coefficients for the following linear underdetermined Diophantine equation 
\begin{eqnarray}\label{eq:dio}
k = \alpha_1 x_1 + \alpha_2 x_2 + \dots \alpha_{r}x_{r},
\end{eqnarray}
where $k = k_{min},k_{min}+1,\dots,k_{max}$ and $r$ denotes the number of eligible hyperstubs -- a node with degree $k=3$ can only go on to form subgraphs where the hyperstubs contain no more than three edges --, for the given degree $k$ and which is solved subject to the constraint $\boldsymbol{x} \in \mathbb{N}_0^{r}$. A solution $\boldsymbol{x}$ of this equation corresponds to the number of each type of hyperstubs required to result in a node of degree $k$. For example, if $\alpha_1$ and $\alpha_2$ take values $1$ and $2$ corresponding to hyperstubs of $G_0$ and $G_\triangle$ respectively and the degree of the node is $k=5$, the Diophantine equation would take the form $5 = x_1 + 2x_2$ and the solution space of this equation is given by the pairs $(x_1,x_2) =\{(5,0),~(3,1),~(1,2)\}$. In general these equations may be solved recursively by fixing a trial value $x_i = j$ and reducing the dimensionality of the equation by absorbing this term. This is repeated until the equation becomes of the standard form: $k' = \alpha_1x_1 + \alpha_2x_2$, which can be solved explicitly. A solution obtained this way will form a single solution of the original equation. This process is then repeated for a different starting trial solution, and since we seek only positive solutions and $k$ is finite, the corresponding solution space has a finite number of elements. Matlab code for this process is available at \url{https://github.com/martinritchie/Network-generation-algorithms}.

Once the entire solution space for each degree has been found it is possible to start forming the hyperstub degree sequences. To proceed, the algorithm works sequentially through the degree sequence $D =\{d_1,d_2, \dots, d_N\}$ of the $N$ nodes, where $d_i \in \{k_{min}, k_{min}+1, \dots, k_{max}\}$. By selecting at random a solution from the solution space that corresponds to $k=d_i$, that specifies the hyperstub configuration, and by concatenating all the selected solutions for all the nodes a hyperstub degree sequence of dimension $h \times N$, where $h$ denotes the total number of hyperstubs induced by the input subgraphs, is formed.

For incomplete subgraphs it is not possible to select solutions of the Diophantine equations' solution spaces at random. The reason for this is two-fold: (1) not all asymmetric subgraphs are composed of equal quantities of each of their constituent hyperstubs, and (2) hyperstubs with lower stub cardinality will appear more frequently than hyperstubs of higher stub cardinality because hyperstubs with fewer edges can be more readily accommodated into the degree of a node. Problem (1) may be addressed by representing every hyperstub induced by a subgraph in the vector of coefficients opposed to grouping hyperstubs by their stub cardinality. Problem (2) may be addressed by decomposing hyperstubs generated in excess into simple/classical edges. This is the approach we take in our implementation. This choice is motivated by its wider applicability. For example, when using $G_\boxslash$ as an input subgraph there will be more degree-2 corners generated than degree-3 corners. However, once all degree-3 corners are allocated to $G_\boxslash$ subgraphs any leftover degree-2 corners may be decomposed back into stubs that can form edges, thus preserving the degree sequence. Finally, it should be noted that if the input subgraphs do not admit hyperstub combinations that can sum to a  particular degree in the network then the proposed method will fail. For example, it is almost always necessary to include $G_0$ (edge) as input.  

Pseudo-code for the UDA algorithm is given in Appendix~\ref{UDA}, and the Matlab code is available from \url{https://github.com/martinritchie/Network-generation-algorithms}.

\subsubsection{A priori clustering calculation}\label{sec:UDA_c}

The global clustering coefficient is defined as the ratio between the total number of triangles and the total number of connected triples of nodes $\triangle + \vee$, since each triangle contains 3 triples of nodes: $C = \frac{\triangle}{\triangle + \vee}$. It should be noted that each unique triangle is counted 6 times and each unique triple is counted twice. The number of triples incident to a node of degree $k$ is given by $\triangle + \vee = k(k-1)$ since a node will form a triple with every pair of its neighbours and each triple is counted twice. The expected number of triples for a node of degree $k$ is therefore given by $P(K=k)\times k(k-1)$, where $P(K=k)$ is the probability of finding a node of degree $k$. The expected number of triangles incident to a node of degree $k$, $\langle \triangle_k \rangle$, may be obtained from the Diophantine equations' solution space associated with that degree. To do this, one needs to sum all occurrences of triangle corners, regardless of which subgraph they belong to, from that solution space and divide by the number of solutions in that particular solution space, since solutions are selected uniformly at random. Finally we are in a position to compute the global clustering coefficient as
\begin{eqnarray}
C = \sum_{k=2}^{k_{max}} \frac{\langle \triangle_k \rangle}{P(K=k) \times k(k-1).}
\end{eqnarray}
For example, let us consider the homogeneous network with $k=5$ and the input subgraphs $G_0$ and $G_\boxslash$. These subgraphs induce the vector of coefficients $\boldsymbol{\alpha} = (1,2,3)$ that, for $k=5$, has the following solution space
\[\begin{array}{rccccc}
         G_0:   & 5 & 3 & 2 & 1 & 0,\\
         g_2:   & 0 & 1 & 0 & 2 & 1, \\
         g_3:   & 0 & 0 & 1 & 0 & 1,
\end{array}\]
where the rows give the number of each hyperstub, the columns give an individual solution and $g_2$ and $g_3$ denote the double and triple hyperstub of $G_\boxslash$ respectively. From this we may calculate the expected number of triangles $\langle \triangle_5 \rangle$. In this example we can see that on average for every $g_3$ corner the UDA algorithm will generate two $g_2$ corners. Since the excess $g_2$ corners will be decomposed into edges, one observes that $g_2$ and $g_3$ will be generated in equal quantities. So the expected number of $g_2$ is given by the expected number of $g_3$, e.g., $2/5$ per node. Since $g_2$ denotes a triangle corner, the number of $g_2$ corners also gives the total 
number of triangles, that is uniquely counted and per node. So the expected number of triangle per node is $12/5$, each triangle being counted 6 times, and this network will have a theoretical global 
clustering of $C=0.12$. Computationally, we verify this by generating such networks with $N=5000$, and find that the number of open triples and triangles is  exactly $|\vee| = 100000$ and $|\triangle| = 
12120$, resulting in a global clustering of $0.1212$, as expected. 

\subsection{Cardinality matching -- CMA}\label{sec:CMA}

The cardinality matching algorithm (CMA) requires as input a degree sequence, a set of subgraphs and corresponding \emph{subgraph sequences}, i.e., multiple sequences specifying to which and how many subgraphs nodes belong to. Note that these sequences are not yet allocated to nodes. The algorithm proceeds to allocate hyperstubs of subgraphs to nodes that have a sufficient number of stubs to accommodate the hyperstub degree. The algorithm outputs hyperstub degree sequences, from which the input degree sequence may be recovered exactly. This then can be used to realise a network based on a modification of the configuration model.

To generate a CMA network one needs to first decide on a degree sequence $D$, a subgraph set $G = \{G_1, G_2, \dots, G_l\}$ and a set of subgraph sequences $S = \{S_1,S_2,\dots, S_l\}$, where $S_j(k)$, with $j=1, 2, \dots, l$ and $k=1, 2,\dots, N$, gives the number of times a node will be part of a $G_j$ subgraph without specifying the precise hyperstubs that connect the node to a $G_j$ subgraph. Our goal is to map the subgraph sequences into hyperstub sequences which can then be allocated to nodes that can accommodate them. From the hyperstub sequence, it is possible to work out the lower bound on the degree of nodes that can accommodate a specific hyperstub sequence. To complete this mapping one needs to differentiate between complete and incomplete subgraphs.

For complete subgraphs the subgraph sequence is identical to its hyperstub sequence since there is only one way or hyperstub by which a node can connect to such a subgraph. Thus, multiplying the hyperstub degree by the number of edges in the hyperstub will give us the lower bound on the degree of nodes that can accommodate the hyperstub sequence. For incomplete subgraphs the subgraph sequence does not specify how the node connects to the subgraph. Hence, we need to determine how the various hyperstubs are allocated to nodes. To see how to do this let us consider an arbitrary subgraph $G$ with subgraph sequence $S$. Given that the subgraph has $m$ distinct hyperstubs, let $p=(p_1, p_2, \dots, p_m)$ be the vector of probabilities of picking different hyperstubs. We note that the values of $p$ reflects the proportion of each hyperstub found in the subgraph. For example, $G_\boxslash$ has two distinct hyperstubs that both appear with multiplicity two, in this case $p=(1/2,1/2)$. This will ensure that their numbers are balanced and subgraphs can be formed.

Next, using the multinomial distribution corresponding to subgraph $G$, $M^{G}(s^{G}_i,P)$ where $s^{G}_i$ denotes the subgraph sequence of index $i$ (this is not yet a node label), we pick hyperstub types to transform the subgraph sequence into hyperstub degree. For each $s^{G}_i$ this will result in a vector of length $m$ specifying the exact number of each hyperstub. It is possible to concatenate all the resulting choices from all multinomial distributions $M^{G}(s^{G}_i,p)$, where $i=1,2, \dots, N$ to form the following matrix
$$
\bordermatrix{
       & s_1^{G}    & s_2^{G} & \dots & s_N^{G} \cr
h_1^{G}    & h_1^{G}(1) & h_1^{G}(2) & \dots & h_1^{G}(N) \cr
h_2^{G}    & h_2^{G}(1) & h_2^{G}(2) & \dots & h_2^{G}(N) \cr
\vdots & \vdots & \ddots & \ddots & \vdots \cr
h_m^{G}    & h_m^{G}(1) & h_m^{G}(2) & \dots & h_m^{G}(N)
\cr}=H^{G},
$$
where $h_i^{G}(j)$ denotes the number of $h_i$ hyperstubs contributing to the subgraph degree $s_j^{G}$. We now need to compute the total number of edges specified by each column of the above matrix or by the hyperstub degree. This is given by $H^{G}(i)=\sum_{j=1}^{m}|h_j^{G}|h_j^{G}(i)$ that denotes the total number of edges required by the subgraph degree $s_i^{G}$, and where $|h_j^{G}|$ represents the number of edges needed to form hyperstub $j$ in subgraph $G$ and $i=(1,2,\dots,N)$. This process needs to be repeated for each subgraph to be included in the networks construction, i.e., for each subgraph $G_i$ with subgraph sequence $S^{G_i}=(s_1^{G_i},s_2^{G_i},\dots, s_N^{G_i})$ there is a corresponding $H^{G_i}$ with elements that the algorithm will use as the lower bound on the degree of the nodes that can accept such a selection of hyperstubs.

The algorithm then proceeds by choosing the largest values, $H_{\max}$, from all $H^{G_i}$ matrices, and this is used as the lower bound on the degree of nodes that can accept the hyperstub configuration associated with $H_{\max}$, i.e., have enough edges of the classical type. From this list of all nodes with degree equal to or larger than $H_{\max}$, a node is selected uniformly at random. The degree of the selected node is reduced accordingly, and the index of the node is now associated with the hyperstub degree to $H_{\max}$. This node is then removed from the pool of eligible nodes for that particular subgraph, as otherwise it may be selected twice for the same subgraph thus violating the subgraph degree sequence. Similarly, the element $H_{\max}$ is also removed from the pool of subgraph degree sequences that have yet to be allocated to nodes. This needs to be repeated until all elements of each subgraph degree sequence are allocated to nodes. Any edges that are not allocated to a particular hyperstub or subgraph are left to form edges.

In some cases it may be necessary to impose some cardinality constraints on the subgraph sequences. Obviously, if the network is homogeneous with $k=3$ we cannot include complete pentagon subgraphs or allocate two $G_\triangle$ subgraphs to each node. More generally, it may be necessary to constrain the moments of the subgraph sequences. Let $\langle k \rangle$ denote the mean degree of the given degree sequence and let $G_i$ be a subgraph composed of a single hyperstub with cardinality $\alpha$ and having subgraph degree sequence with mean $\langle s \rangle$ then: $ \langle \alpha s \rangle = \alpha \langle  s \rangle \leq \langle k \rangle$ is a necessary condition for the two sequences to be graphical. In the case of more than one hyperstub, this is extended to $\sum_{i = 1}^m\alpha_i \langle s_i \rangle \leq \langle k \rangle$, where $m$, $\alpha_i$ and $s_i$  denote the number of hyperstubs, hyperstub cardinality and associated subgraph sequence respectively. For the networks generated in this paper, the degree sequence and subgraph sequences were measured from networks previously generated by the UDA such that prior knowledge about the sequences being graphical was available without the need to impose any such constraints.  

Clustering calculations for this algorithm are trivial since the subgraph degree sequences are known. One simply sums a sequence and then multiplies this figure by the number of triangles induced by that subgraph, being careful not to double count across multiple sequences for the same subgraph. The number of triples of connected nodes can be calculated following the method given for the UDA given in Section~\ref{sec:UDA_c}. Pseudo-code for the CMA is given in Appendix~\ref{CMA}, with the corresponding Matlab code available from \url{https://github.com/martinritchie/Network-generation-algorithms}.
 
\subsection{Connection process}\label{sec:CP}

We describe this process for a single incomplete subgraph. The case of the complete subgraph is trivial and has already been described (see Section~\ref{sec:intro}). This process was first presented by Karrer \& Newman \cite{karrer2010random}. Consider a subgraph composed of three different hyperstub types, $h_1$, $h_2$ and $h_3$ that occur with a multiplicity of 1, 2 and 3 respectively, i.e., the subgraph is composed of 6 nodes. For these hyperstub sequences to be graphical we require
\begin{eqnarray}
\sum_{i=1}^N |h_1|_i = \frac{1}{2}\sum_{i=1}^N |h_2|_i = \frac{1}{3}\sum_{i=1}^N |h_3|_i,
\end{eqnarray}
where $|h_i|_j$ specifies the $h_i$ hyperstub degree of node $j$. If these conditions are not met, one needs to decompose any surplus hyperstubs into stubs that may form classical edges in order to preserve the degree sequence. 

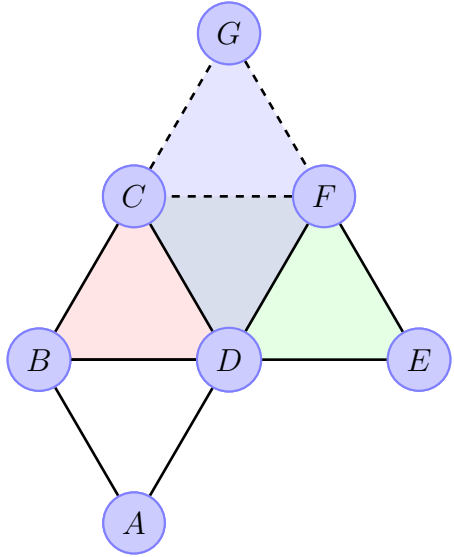
\begin{figure}[!htbp]
\begin{center}
\begin{tikzpicture}[scale=.5]
           \node[place, minimum size = 8mm] (A) at ( 2.5, -4.33) {\scalebox{1}{$A$}};
           \node[place, minimum size = 8mm] (B) at ( 0,0) {\scalebox{1}{$B$}};
           \node[place, minimum size = 8mm] (D) at ( 5,0) {\scalebox{1}{$D$}};            
           \node[place, minimum size = 8mm] (E) at ( 10,0) {\scalebox{1}{$E$}};
           \node[place, minimum size = 8mm] (C) at ( 2.5,4.33) {\scalebox{1}{$C$}};
           \node[place, minimum size = 8mm] (F) at ( 7.5, 4.33) {\scalebox{1}{$F$}};            
           \node[place, minimum size = 8mm] (G) at ( 5, 8.65) {\scalebox{1}{$G$}};

           \path[fill=red!20,opacity=.5] (B.center) to (C.center) to (F.center) to (D.center) to (B.center);
           \path[fill=green!20,opacity=.5] (C.center) to (F.center) to (E.center) to (D.center) to (C.center);
           \path[fill=blue!20,opacity=.5] (C.center) to (G.center) to (F.center) to (D.center) to (C.center);

           \node[place, minimum size = 8mm] (A) at ( 2.5, -4.33) {\scalebox{1}{$A$}};
           \node[place, minimum size = 8mm] (B) at ( 0,0) {\scalebox{1}{$B$}};
           \node[place, minimum size = 8mm] (D) at ( 5,0) {\scalebox{1}{$D$}};            
           \node[place, minimum size = 8mm] (E) at ( 10,0) {\scalebox{1}{$E$}};
           \node[place, minimum size = 8mm] (C) at ( 2.5,4.33) {\scalebox{1}{$C$}};
           \node[place, minimum size = 8mm] (F) at ( 7.5, 4.33) {\scalebox{1}{$F$}};            
           \node[place, minimum size = 8mm] (G) at ( 5, 8.65) {\scalebox{1}{$G$}};

          %%%%%%%%%%%%%%%%%%%%%%
           \draw [line width=1pt, dashed] (C) -- (G);
           \draw [line width=1pt, dashed] (G) -- (F);
           \draw [line width=1pt, dashed] (F) -- (C);

           \draw [line width=1pt] (E) -- (F);
           \draw [line width=1pt] (E) -- (D);
           \draw [line width=1pt] (D) -- (F);

           \draw [line width=1pt] (C) -- (B);
           \draw [line width=1pt] (D) -- (C);
           \draw [line width=1pt] (B) -- (D);

           \draw [line width=1pt] (A) -- (B);
           \draw [line width=1pt] (A) -- (D);
           \draw [line width=1pt] (B) -- (D);
\end{tikzpicture}
\end{center}
\caption{Unintended generation of subgraphs with overlap. Despite satisfying the generation constraints given in Section~\ref{sec:CP}, the addition of triangle (C,G,F) to toast (A,B,C,D) and triangle (D,F,E) results in 3 unintended distinct toasts \{(B,C,F,D) in red, (D,C,F,E) in green, and (D,C,G,F) in blue\} overlapping on one unintended triangle (C,F,D), in gray.}
\label{fig:byproductexample}
\end{figure}

Using the hyperstub sequences, one can create three dynamic lists for the three hyperstub types where a node appears with multiplicity equal to its hyperstub degree. Once the dynamic lists are fully populated, the connections process can start. This is done by selecting the following: 1 node from the $h_1$ bin, 2 from the $h_2$ bin and 3 from the $h_3$ bin, and all the selection processes done  uniformly at random and without replacement. Before forming the connections between these 6 nodes, one must ensure  that: (1) the selection contains no duplicates (that will form self-edges) and (2) that no single pair of nodes are already connected. If a connection already exists, a multi-edge may form and/or subgraphs will share edges. If neither of these conditions are violated then the connections may be formed. Otherwise all nodes are returned to their bins and a new selection is made. As previously discussed, it is possible to delete self and multi-edges, however, this will destroy the degree sequence. The method of reselecting nodes has been previously introduced and is known as the \emph{matching algorithm} \cite{milo2003uniform}. It is possible that after many selections no valid combinations of nodes remain. For example, all bins may contain the same node. In this and other non-viable cases, all bins are re-populated and the connection process is started anew. It should be noted that, as none of the construction constraints discussed above involve the neighbours of the nodes being connected, it is possible for previously created subgraphs to become connected into a set of subgraphs with overlap, see Figure~\ref{fig:byproductexample} for an illustration. Evidence of such by-products will be shown in Section~\ref{sec:UDA_CMA}.

\subsection{The Big-V algorithm}\label{sec:bigv}
The Big-V algorithm does not generate networks as such, but is a widely-used, see~\cite{house2010impact,house2011insights,ritchie2014higher,green2010large} for example, degree-preserving rewiring algorithm making it possible to control clustering. At each iteration, the algorithm selects a linear chain of 5 nodes at random, e.g., $\{a,b,c,d,e\}$ with 4 edges $\{(a,b),(b,c),(c,d),(d,e)\}$. It then delete edges $(a,b)$ and $(d,e)$ to form $(a,e)$ and $(b,d)$. When starting from an unclustered network, this process will lead to at least one extra $G_\triangle$ being created~\cite{bansal2009exploring}. This is repeated until the desired level of clustering is achieved. It is possible to include a Metropolis-style augmentation whereby at each step the local clustering coefficient is computed for the five nodes before and after rewiring, and the rewired configuration is only accepted if it results in an increase in average local clustering. It is worth noting that this algorithm leads to a positive degree-degree correlation which was not necessarily present in the original network.

In this paper, we use the Big-V algorithm to demonstrate that our newly proposed algorithms are able to sample from a larger part of the state space of all possible networks with a given degree sequence and global clustering coefficient.

\subsection{Models of contagion}\label{sec:contagion}

In order to illustrate the impact of network structure -- and higher-order structure particularly -- different epidemic dynamics were simulated on the generated networks. Three different models were chosen: Susceptible-Infected-Susceptible ($SIS$), Susceptible-Infected-Recovered ($SIS$) 
and complex contagion \cite{miller2015complex, osullivan2015mathematical}. To simulate $SIS$ and $SIR$ dynamics, the fully susceptible network of nodes is perturbed by infecting a small number of nodes. Infected nodes spread the infection to susceptible neighbours at a per-link rate of infection $\tau$. Infected individuals recover independently of the network at rate $\gamma$ and become  susceptible again (for $SIS$ dynamics) or become removed (for $SIR$ epidemics). In contrast to the infection process in the previous two dynamics, the complex contagion process requires that susceptible nodes are exposed to multiple infectious events before becoming infected. These events must be from different infectious neighbours as only the first infection attempt from an infectious node counts.  This critical infection threshold for each node is set in advance and is usually bounded from above by the degree of the node. To simulate the complex contagion dynamics, nodes are allocated infection thresholds $r_i\in \mathbb{N}$, where $i=1,2\dots, N$, and the fully susceptible population of nodes is perturbed by infecting an initial number of nodes chosen at random. In this model a susceptible node $i$ becomes infected as soon as it has received at least $r_{i}$ infectious contacts from $r_{i}$ distinct infectious neighbours. There is no recovery in this model and infected individuals remain infected for the duration of the epidemic. 

\section{Results}\label{sec:results}

\subsection{Algorithm validation}\label{sec:validation}

To validate our algorithms, we generated a number of networks with pre-specified degree distribution and subgraph set, as well as a multinomial distribution of subgraph corners or hyperstubs around nodes. We verified that the networks generated were as expected given the input. 

\begin{figure}[!htbp]
\centering
\begin{subfigure}{0.45\textwidth}
  \centering
  \includegraphics[width=1\linewidth]{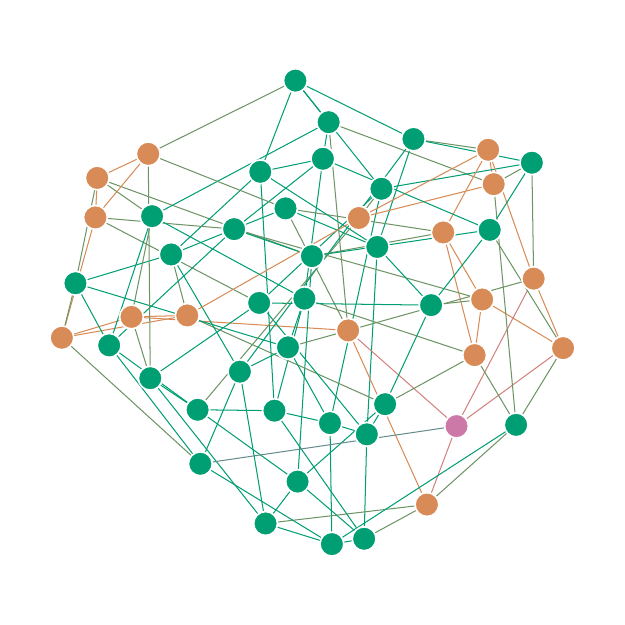}
  \caption{Random}
  \label{fig:small_random}
\end{subfigure}%
\begin{subfigure}{0.45\textwidth}
  \centering
  \includegraphics[width=1\linewidth]{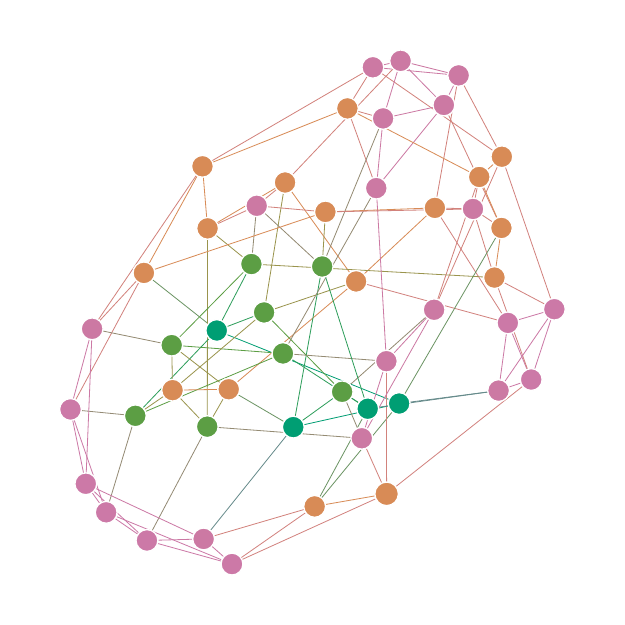}
  \caption{Big-V, C=0.22}
  \label{fig:small_big_v}
\end{subfigure}
\vskip\baselineskip
\begin{subfigure}{0.45\textwidth}
  \centering
  \includegraphics[width=1\linewidth]{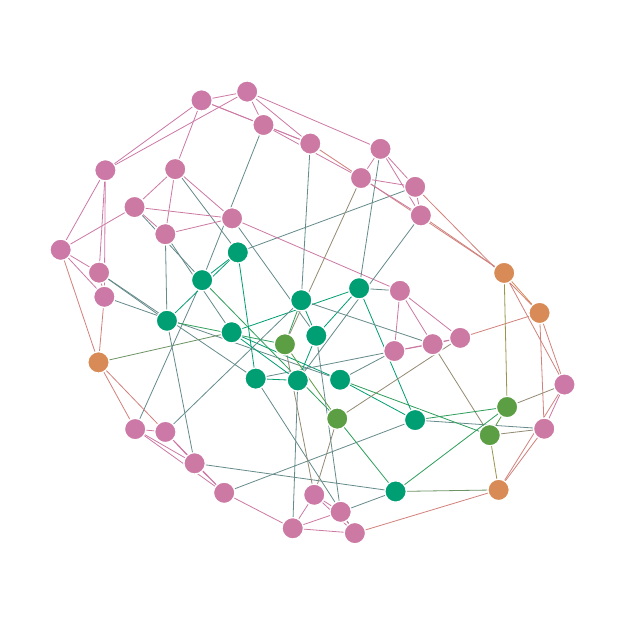}
  \caption{UDA, C=0.22}
  \label{fig:small_uda}
\end{subfigure}%
\begin{subfigure}{0.45\textwidth}
  \centering
  \includegraphics[width=0.9\linewidth]{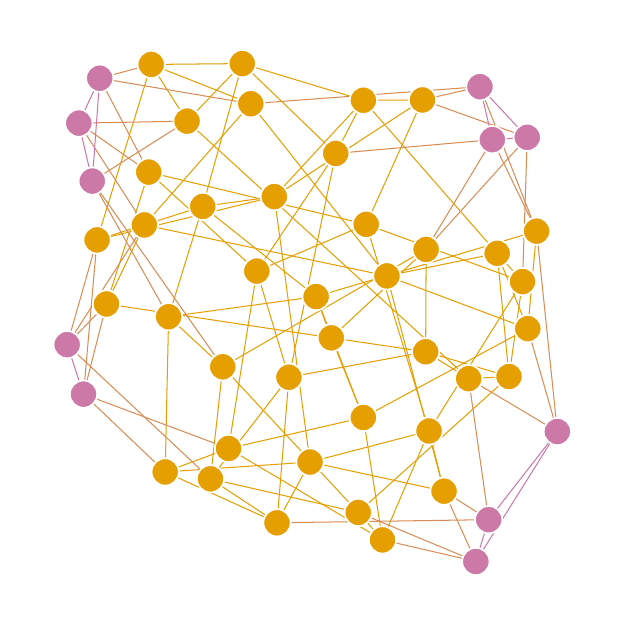}
  \caption{CMA, C=0.22}
  \label{fig:small_card}
\end{subfigure}
\caption{Small networks generated by the Big-V, UDA and CMA algorithms. All networks have the same homogeneous degree sequence with $k=5$. The Big-V algorithm re-wired the random network, Figure~\ref{fig:small_random}. The UDA was parametrised with subgraphs $G_0$ $G_\boxslash$ and $G_\boxtimes$. The CMA was parametrised so that every node was incident to 2 $G_\triangle$. The Big-V, UDA, and CMA networks all have a global clustering coefficient of $C=0.22$. The network nodes are coloured so that green/orange/pink denotes nodes of low/medium/high clustering, respectively.}
\label{fig:example_nets}
\end{figure}

As described in Section~\ref{sec:MM} the algorithms preserve the degree sequence, permitting at most a single edge to be deleted if the degree sequence sums to an odd number. The ability to exercise control over the networks' subgraph topology is illustrated by Figure~\ref{fig:example_nets}. Note that Figure~\ref{fig:small_random} shows a \emph{random} network that includes $G_\triangle$ subgraphs. When constructing networks using the configuration model it is possible to create $G_\triangle$ subgraphs with non-zero probability and this is to be expected~\cite{newman2009networks}. However, this is a function of mean degree not network size, and this probability goes to zero with network size going to infinity. 

\begin{table}[!htbp]
\centering
\begin{tabular}{ccccccccc}\rowcolor{LightBlue} 
       & c4 & d4 & e4 & i4 & s4 & t3 & u3 & u4 \\ \hline
Random & 0 & 0 & 42 & 17 & 446 & 6 & 482 & 1706 \\\rowcolor{LightBlue} 
Big-V  & 1 & 23 & 10 & 10 & 212 & 7 & 386 & 1220 \\
UDA    & 7 & 10 & 22 & 5 & 243 & 1 & 389 & 1239 \\\rowcolor{LightBlue} 
CMA   & 0 & 9  & 10 & 40 & 185 & 24 & 389 & 1201
\end{tabular}
\caption{Subgraph counts for the networks of Figure~\ref{fig:example_nets}. Note: if one adds a single $G_\triangle$ so that it shares a single edge with a $G_\boxslash$ and this edge is not the diagonal edge of $G_\boxslash$, then $d4$ increases by one but $t3$ will have only increased by one, not two. We note that $2\cdot d3$ yields the maximum number of possible $G_\triangle$ induced by $G_\boxslash$. In general, calculating the number of $G_\triangle$ in this way will always yield the maximum possible count but not necessarily the true count because a single $G_\triangle$ could be shared by more than one $G_\boxslash$.}
\label{tab:example_counts}
\end{table}

%-------------------------------------------------------------
% Algorithm validation.
%-------------------------------------------------------------
% example networks

To properly demonstrate the proposed algorithms' control over the building blocks in the network, we used a recently described subgraph counting algorithm~\cite{ritchie2014higher} to count the number of subgraphs \emph{a posteriori}. In our implementation we counted subgraphs composed of 4 nodes or less -- see the top two rows of Figure~\ref{fig:examples}, as well as 5- and 6-cycles. Table~\ref{tab:example_counts} provides the subgraph counts for the networks displayed in Figure~\ref{fig:example_nets}. It confirms that the random network given in Figure~\ref{fig:small_random} contains 6 $G_\triangle$, counted uniquely, as observed above. The table also reveals that, through increasing the frequency of $G_\triangle$, the Big-V algorithm also introduced $G_\boxslash$ and $G_\boxtimes$ subgraphs. The UDA was parametrised with $\{G_0, G_\boxslash, G_\boxtimes\}$ and the table confirms a significant presence of these subgraphs when compared to the random network. Although the CMA was parametrised solely with $G_\triangle$ subgraphs distributed so that each node was incident to 2 $G_\triangle$ subgraphs, the subgraph counts reveal that this network contains 9 $G_\boxslash$ subgraphs. This is a consequence of attempting to generate \emph{small} networks with such a high prevalence of triangles: it is highly likely that the algorithms will select nodes that already share one other common neighbour later in the connection process. One expects the proportion of these events to become increasingly negligible with greater network size.   

\begin{figure}[!htbp]
\centering
\begin{subfigure}{0.49\textwidth}
  \centering
  \includegraphics[width=1\linewidth]{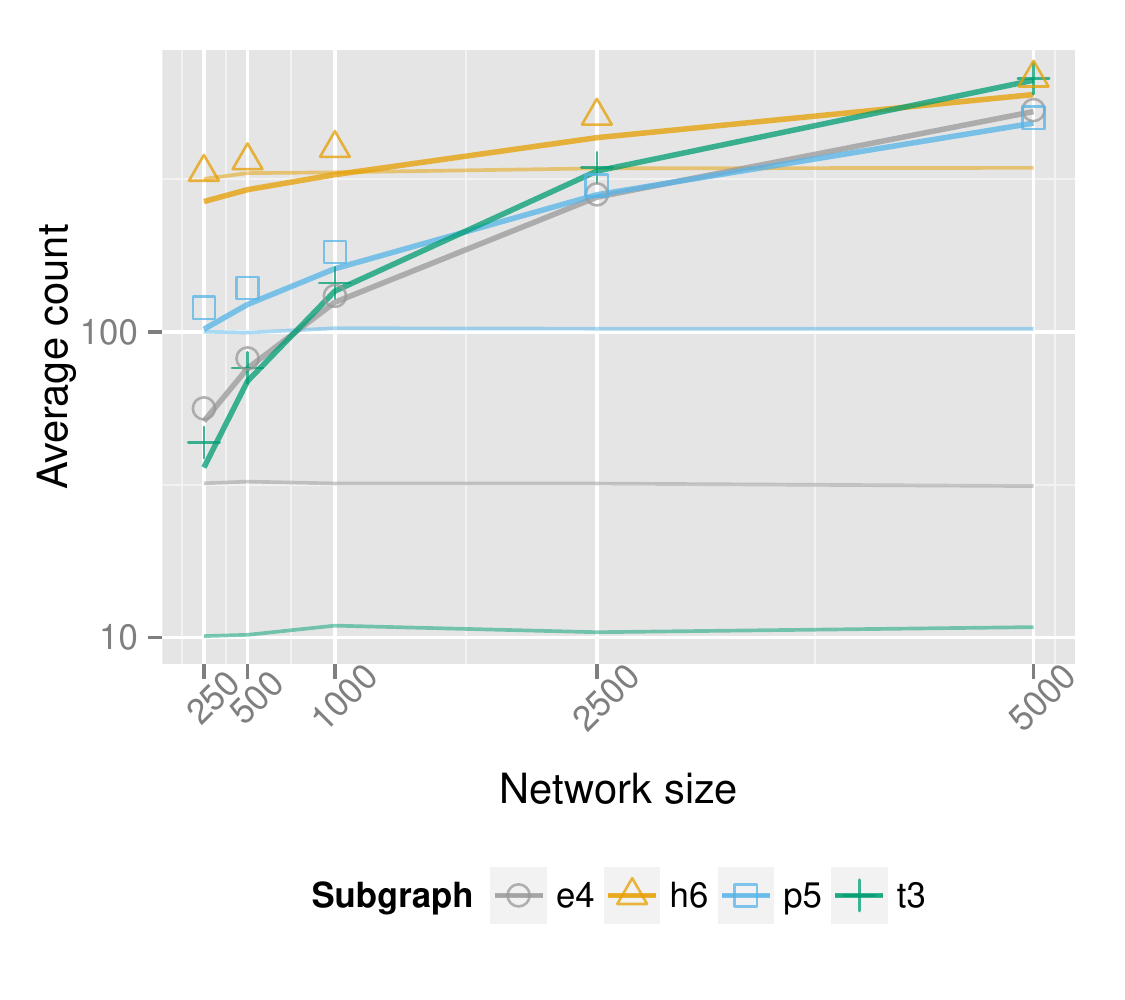}
  \caption{UDA}
  \label{fig:rand_and_uda}
\end{subfigure}%
\begin{subfigure}{0.49\textwidth}
  \centering
  \includegraphics[width=1\linewidth]{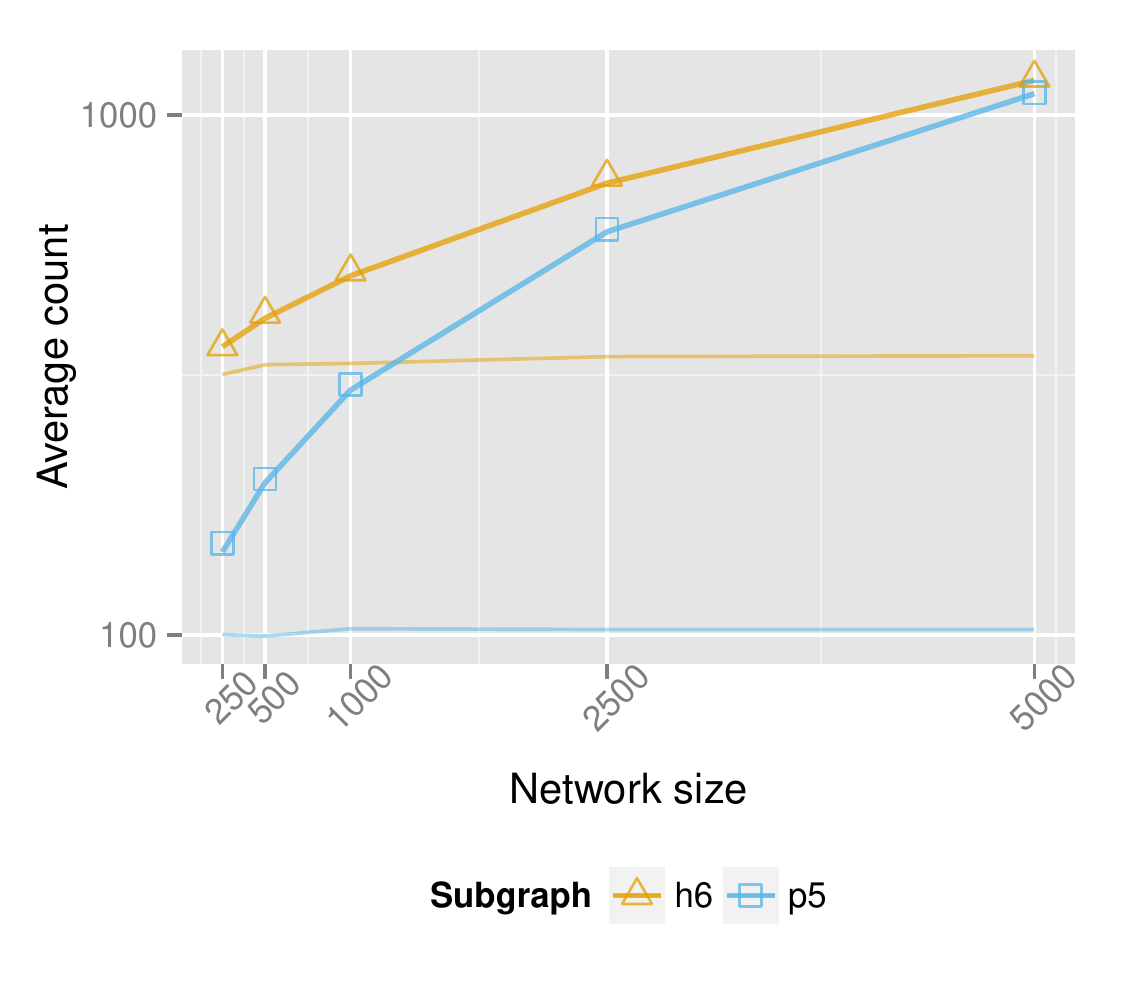}
  \caption{CMA}
  \label{fig:rand_and_card}
\end{subfigure}
\caption{A comparison of subgraphs found in the UDA and CMA networks to their random network analogues and expected counts plotted with thick lines, thin lines and discrete markers respectively. $p5$ and $h6$ denote the counts of $G_{\pentagon}$ and $G_{\hexagon}$ respectively. All networks have the same homogeneous degree sequence with $k=5$ but with increasing size: $N=250,500,1000,2500,5000$, where 100 of each size was generated. (a) The UDA algorithm was parametrised with subgraphs $\{G_\triangle, G_\boxempty, G_{\pentagon}, G_{\hexagon}\}$, and the resulting average subgraph counts are shown on the left. (b) The CMA algorithm was parametrised so that each node was incident to a single $G_{\pentagon}$ and $G_{\hexagon}$ subgraph, and the resulting average subgraph counts are shown on the right. The expected values were calculated by summing the total counts from the subgraph sequences, dividing them by the subgraphs' node cardinality, and adding these figures to the number of subgraphs found as by-products in the random networks.}
\label{fig:homo_k5_counting}
\end{figure}

Next, we used the above motif counting algorithm to evaluate the extent to which the proposed algorithms can exert control over the prevalence of subgraphs in the generated networks. Figure~\ref{fig:homo_k5_counting} compares \emph{measured} counts of subgraphs in UDA and CMA networks with \emph{expected} counts. Here, an important observation must be made at the outset. Even in random networks, cycles ($G_\boxempty$, $G_{\pentagon}$ and $G_{\hexagon}$) appear in significant quantities: 33, 100 and 333 times respectively, and regardless of network size. They are a natural consequence of the fact that the probability of selecting two nodes in different branches of a finite tree-like network is non-zero. Therefore, our \emph{expected} counts are the sum of the counts expected \emph{by construction} and those \emph{measured} in the random networks. For example, since the CMA networks were generated with each node being incident to a single $G_{\hexagon}$ subgraph, a total of 833 uniquely counted $G_{\hexagon}$ subgraphs were expected \emph{by construction} in networks of size $N=5000$. However, because an average of 344 $G_{\hexagon}$ subgraphs were counted in random networks of size $N=5000$, our \emph{expected} count was $833+344=1177$. The \emph{measured} count was found to be 1165. More generally, we found the \emph{expected} counts to match well with the \emph{measured} counts, indicating that the generating algorithms did not create by-products in addition to those observed at random\footnote{Although we will show in Section~\ref{sec:UDA_CMA} that for specific parameterisations of CMA, by-products are possible.}. However, these results also suggest that the level of control exerted by the algorithms over subgraph prevalence depends on how often those subgraphs appear naturally as by-products. Control is strongest for subgraphs that do not appear naturally as by-products. When considering subgraphs that appear naturally with high frequency, e.g., $G_{\pentagon}$, real control over their prevalence can only be achieved if an even higher frequency is imposed, which may not always be possible for a given degree sequence and global clustering. 

% Triangle by-products

In what follows, we set out to highlight differences between the new algorithms compared to classic ones and also to emphasise the diversity within networks generated by the same algorithms.

\subsection{Sampling from a different area of the network state space}\label{sec:UDA_Big_V}

In this section, we seek to highlight the versatility of the proposed generation mechanisms by showing that, given a degree distribution and a global clustering, they sample different areas of the network state space than existing methods such as Big-V. We begin by reminding the reader that the Big-V algorithm searches for paths of 5 nodes and rewires such paths so that additional triangles are created. In other words, the principal building block of this algorithm is the $G_\triangle$ subgraph and subgraphs that may be constructed by overlapping $G_\triangle$ subgraphs. It follows that this algorithm is unlikely to give rise to a higher than expected at random number of $G_\boxempty$ or other `empty' cycles. The UDA algorithm was therefore parametrised with  subgraph family $\{G_0, G_\triangle, G_\boxempty, G_{\pentagon}, G_{\hexagon} \}$. In order to eliminate the effect of degree heterogeneity, a homogeneous degree sequence with $k=5$ was used. The resulting networks had a global clustering coefficient of $C=0.04$, induced by 666 (uniquely counted) $G_\triangle$ subgraphs. We then used the Big-V algorithm to rewire random networks constructed using the same degree sequence until the desired level of clustering, $C=0.04$, was achieved. Significant differences between generated networks would confirm that the Big-V and UDA generated networks are sampled from different areas of the state space of networks satisfying that degree sequence and global clustering. As a further point of reference, data taken from a random network realisation of the degree sequence was included in all of our analyses. Henceforth we shall refer to these three types of networks as network family \textbf{A}.

%-------------------------------------------------------------
% Network Models II - metrics.
%-------------------------------------------------------------

\begin{figure}[!htbp]
\centering
\begin{subfigure}{0.31\textwidth}
  \centering
  \includegraphics[width=1\linewidth]{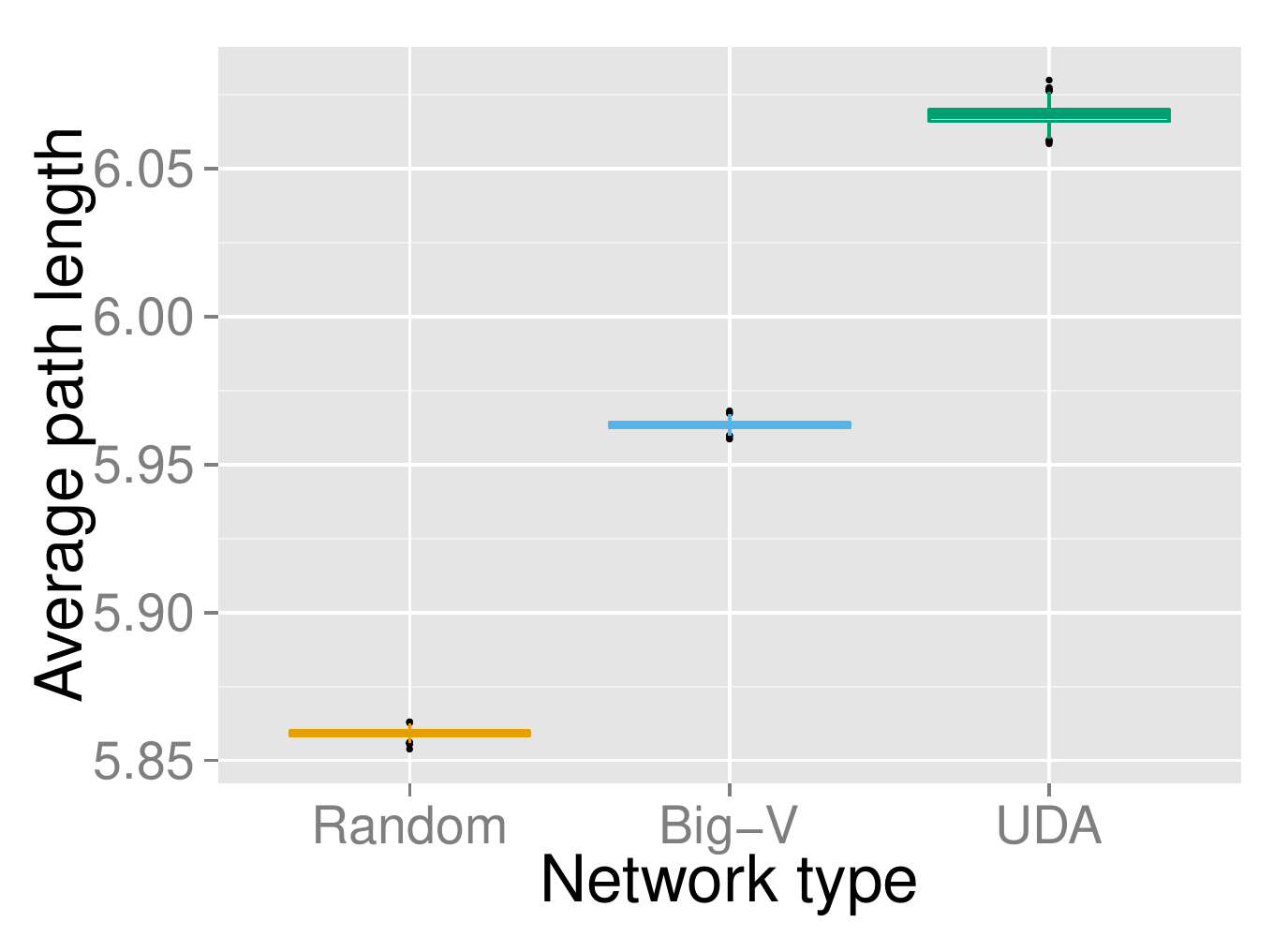}
  \caption{Average path length}
  \label{fig:homo_k5_cycles_avpl}
\end{subfigure}%
\begin{subfigure}{0.31\textwidth}
  \centering
  \includegraphics[width=1\linewidth]{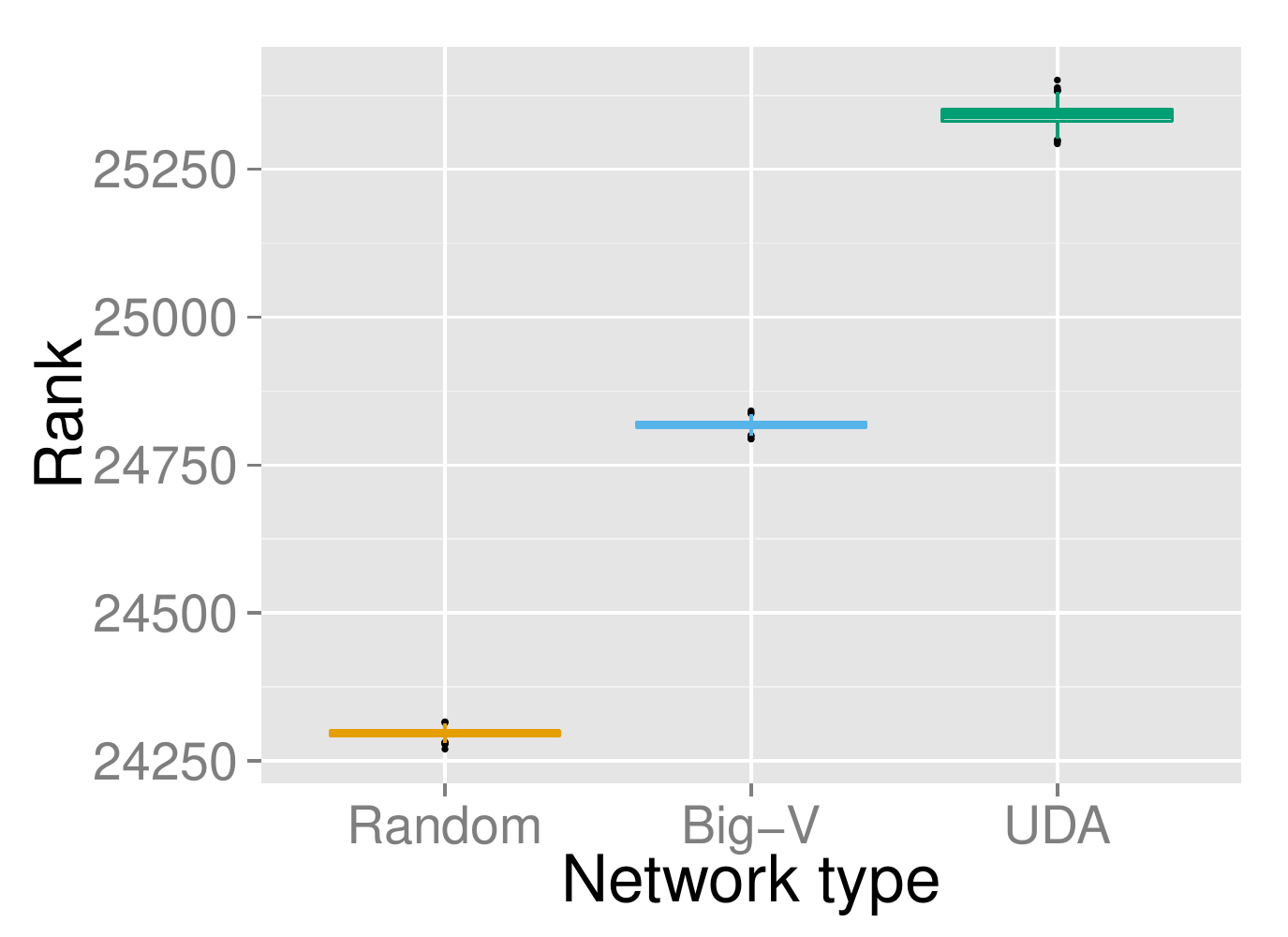}
  \caption{Average betweenness}
  \label{fig:homo_k5_cycles_avg_betw}
\end{subfigure}
\begin{subfigure}{0.31\textwidth}
  \centering
  \includegraphics[width=1\linewidth]{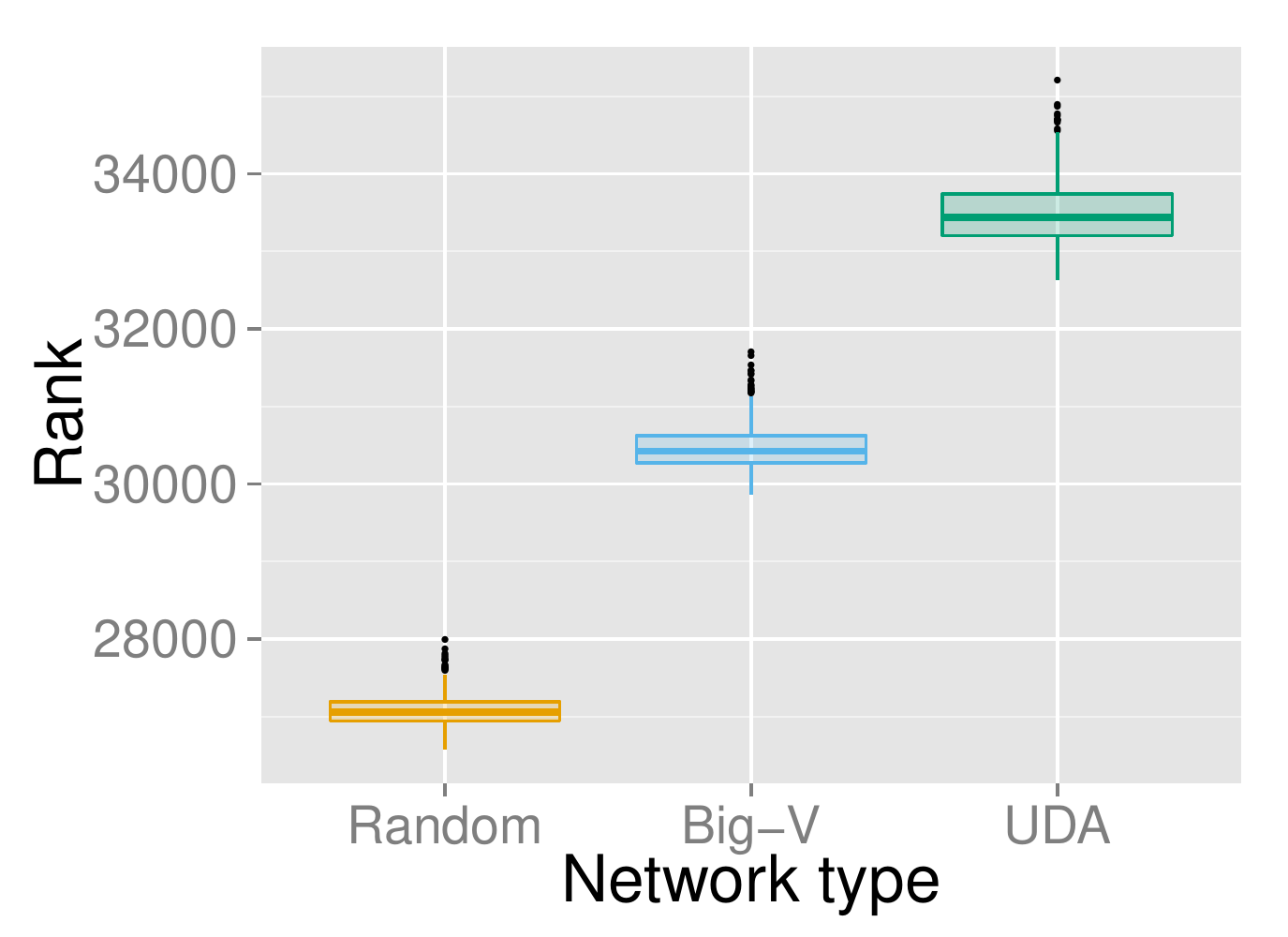}
  \caption{Maximum betweenness}
  \label{fig:homo_k5_cycles_max_betw}
\end{subfigure}
\caption{Plots of the average path length and diameter for homogeneous networks ($N=5000$ and $k=5$) for network family \textbf{A}. The Big-V algorithm was parametrised solely by clustering, in this case $C=0.04$, to best suit the networks produced by the UDA. The differences in average path length, average betweenness centrality and maximum betweenness centrality between the random network and its Big-V analogue were of similar magnitude as the differences between the Big-V network and the cycle-based UDA networks, and these were significant.}
\label{fig:homo_k5_cycles_betw}
\end{figure}

\begin{figure}[!htbp]
\centering
\includegraphics[scale=1]{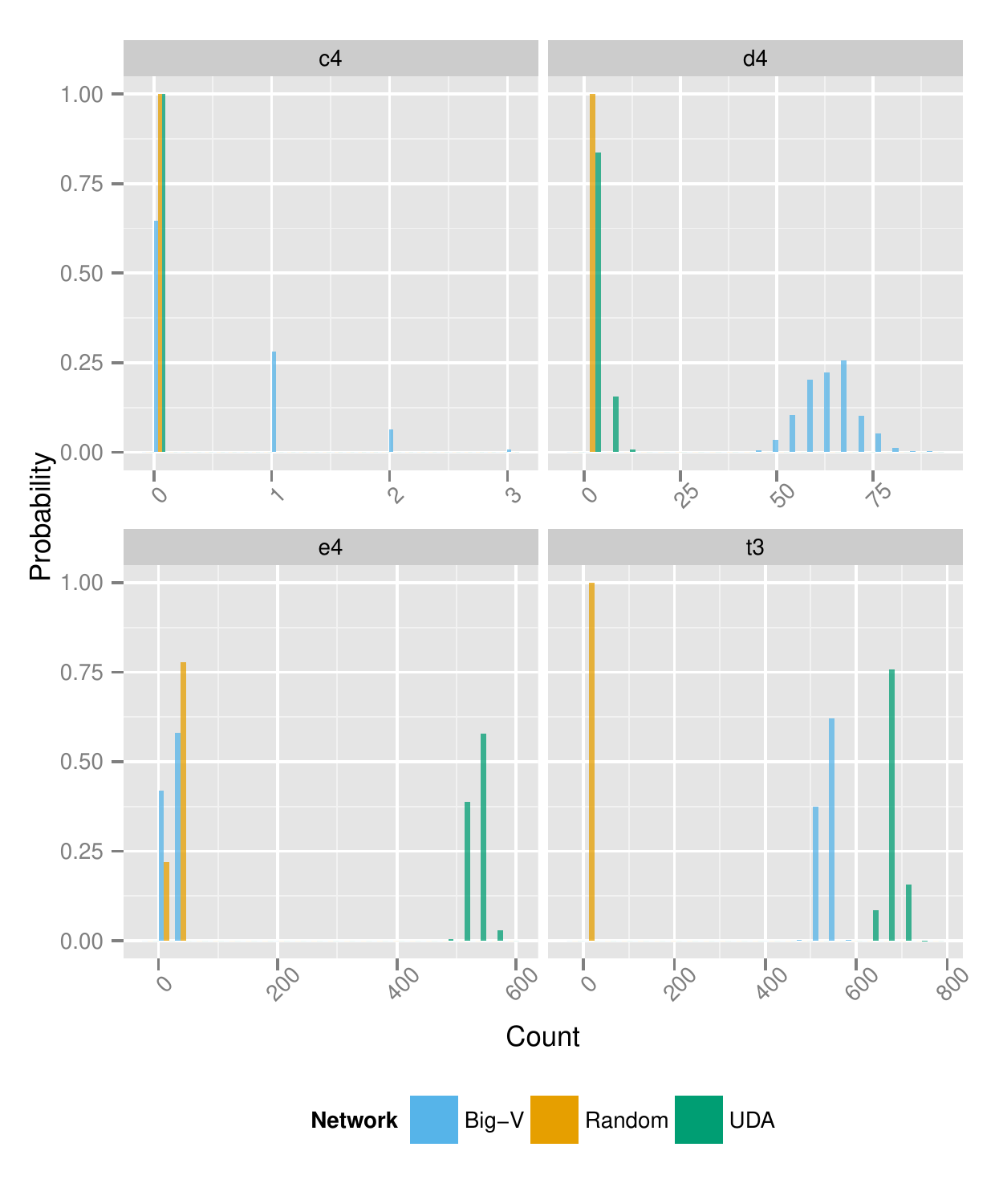}
\caption{Distributions of total number of subgraphs in network family \textbf{A} ($N=5000$, $k=5$). The Big-V and UDA networks have a global clustering coefficient of $C=0.04$. All given counts are unique. The $t3$ counts denote the number of $G_\triangle$ subgraphs that are not involved in any subgraphs of four nodes (i.e., $G_\boxslash$ and $G_\boxtimes$). However, the $c4$ and $d4$ counts may include $G_\triangle$ subgraphs shared by $G_\boxslash$ and $G_\boxtimes$. The number of $G_\boxempty$ subgraphs generated by the Big-V algorithm is very close to the counts found in random networks.}   
\label{fig:homo_k5_counts}
\end{figure}

In Figure~\ref{fig:homo_k5_cycles_betw}, the distributions of the average path length, average betweenness centrality and maximum betweenness centrality for the above networks are given. In general, an increase in clustering results in a higher value of the average path length -- see the average path length of random and Big-V networks in Figure~\ref{fig:homo_k5_cycles_avpl}. This is a known result ~\cite{bansal2009exploring}. Surprisingly, a similar magnitude of difference in average path length and average and maximum  betweenness centrality is observed between the Big-V and UDA networks despite them having the same global clustering, see Figure~\ref{fig:homo_k5_cycles_avpl}, ~\ref{fig:homo_k5_cycles_avg_betw} and ~\ref{fig:homo_k5_cycles_max_betw}, respectively. Output from the subgraph counting algorithm (Figure~\ref{fig:homo_k5_counts}) confirms that, as expected, the Big-V algorithm does not generate more $G_\boxempty$ subgraphs than are observed in the random network. More generally, the results show that the Big-V and UDA networks exhibit markedly different subgraph topologies with the Big-V networks relying heavily on $G_\boxslash$ to cluster the networks unlike UDA networks that rely almost exclusively on $G_\triangle$ not appearing as part of any other subgraph. It may be that such variation was facilitated by the low level of clustering considered, and that with higher clustering, eliciting such differences might be more challenging. However, these results provide evidence that the UDA algorithm can sample from a different part of the state space than the Big-V algorithm.

\subsection{Diversity within the newly proposed algorithms}\label{sec:UDA_CMA}

In this section, we illustrate the diversity of networks generated with UDA and CMA by exploring the impact of subgraph distribution over nodes (for identical degree distribution and global clustering) and how it may change network characteristics. 

To do this we first parametrised the UDA with subgraph family $\{G_0, G_\triangle, G_\boxempty, G_\boxslash, G_\boxtimes\}$ (chosen due to its frequent use in the literature, e.g., ~\cite{ritchie2014beyond, ritchie2014higher, house2010generalised, house2009motif, bansal2009exploring, karrer2010random}), and a heterogeneous degree sequence generated using the Poisson distribution with $\lambda = 5$. Since it is difficult to control for the number of subgraphs that appear in a network generated using the UDA we counted the total number of each subgraph, from UDA-produced subgraph sequences, and used these counts to create alternative subgraph sequences as input to the CMA, see Section~\ref{sec:CMA}, rather than drawing such sequences from a theoretical distribution. The resulting networks were therefore  expected to have identical degree sequence, global clustering of 0.13 and subgraph counts. Since the CMA allows us to choose arbitrary sequences of subgraphs, we opted to push the clustered subgraphs, $\{ G_\triangle, G_\boxslash, G_\boxtimes\}$, onto the higher-degree nodes to accentuate the effect of clustering. We did this by specifying that these subgraphs had to appear with multiplicity greater than one. For example, a degree-three $G_\boxtimes$ hyperstub required a minimum $k=9$-degree node. As previously, we included a random network realisation of the heterogeneous degree sequence for comparison. Henceforth, we shall refer to these three types of networks as network family \textbf{B}. 

%-------------------------------------------------------------
% Network Models III - metrics.
%-------------------------------------------------------------

\begin{figure}[!htbp]
\centering
\begin{subfigure}{0.45\textwidth}
  \centering
  \includegraphics[width=1\linewidth]{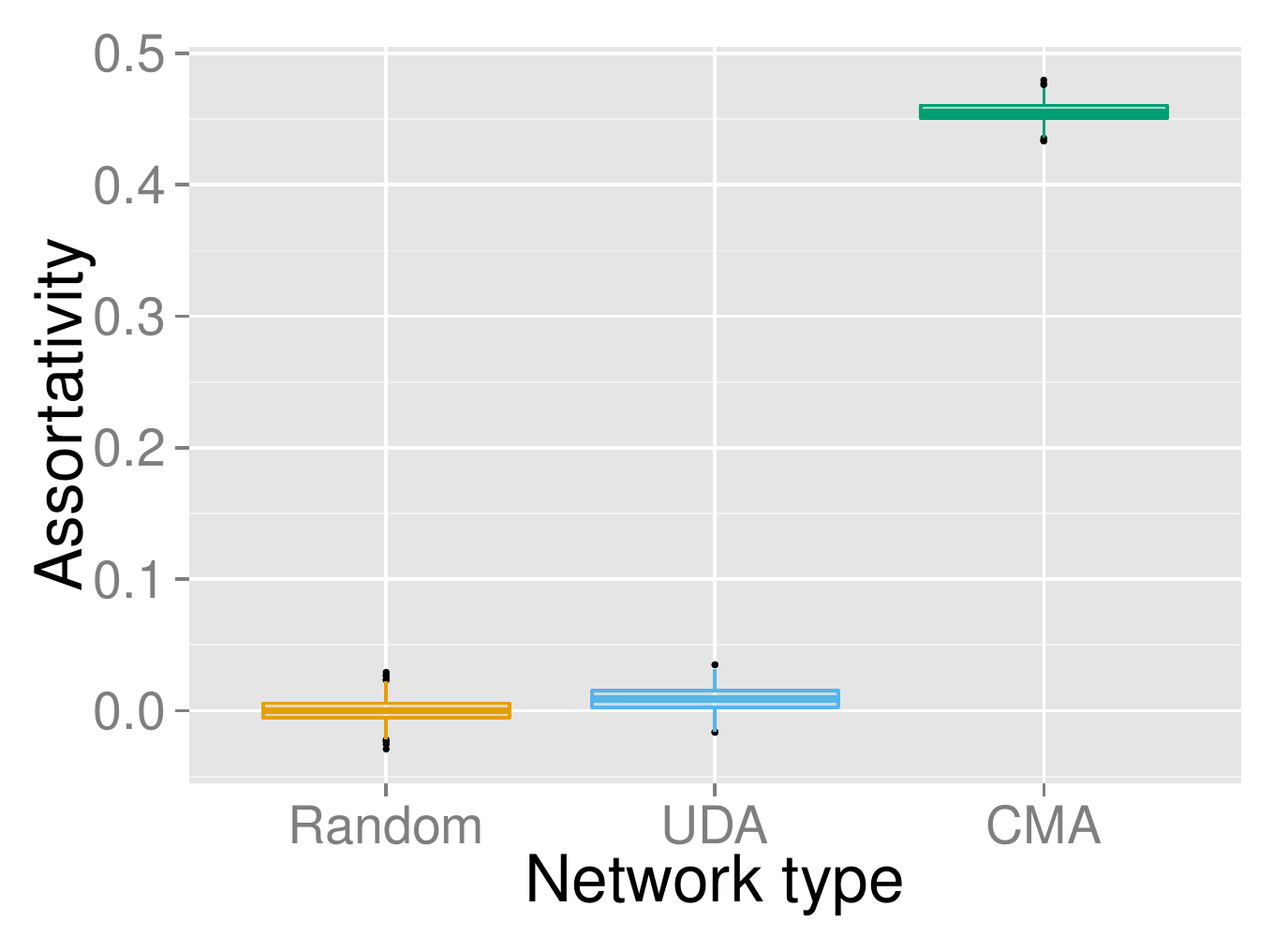}
  \caption{Assortativity}
  \label{fig:pois_k5_ass}
\end{subfigure}%
\begin{subfigure}{0.45\textwidth}
  \centering
  \includegraphics[width=1\linewidth]{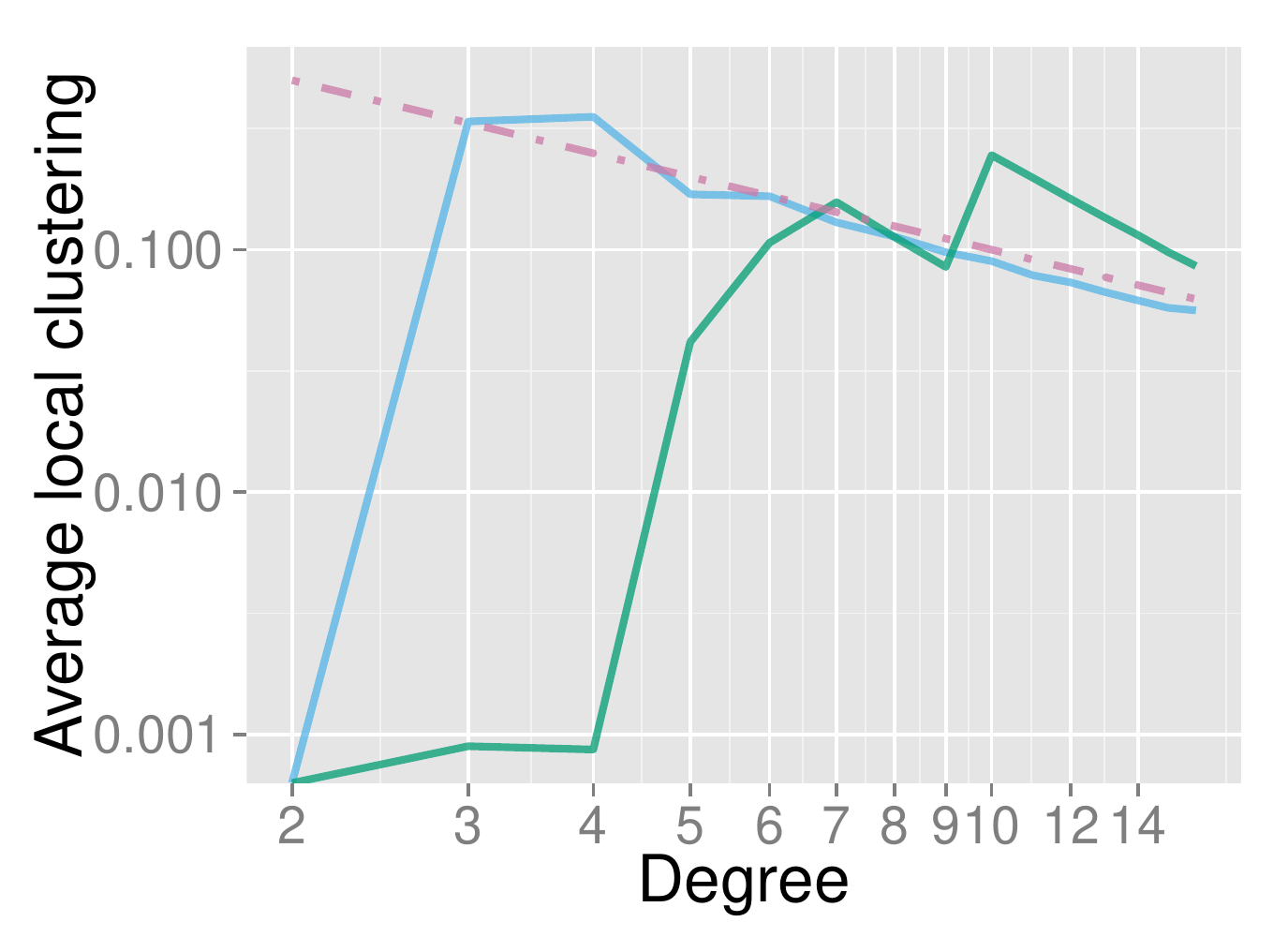}
  \caption{Degree dependent clustering}
  \label{fig:pois_k5_ddc}
\end{subfigure}
\caption{Plots of assortativity and degree-dependent clustering for network family \textbf{B} with $k \sim Pois(5)$. The UDA and CMA networks have a global clustering coefficient of $C=0.13$. The distribution of subgraphs in CMA networks was manipulated so that the clustered subgraphs $\{G_\triangle, G_\boxslash, G_\boxtimes\}$ appeared around nodes with multiplicity greater than one. In order to preserve the subgraph degree sequence these aggregated subgraphs were allocated to the higher degree nodes, resulting in higher assortativity and a more positively skewed distribution of degree-dependent clustering. The dash-dotted line corresponds to $c(k)=k^{-1}$.}
\label{fig:pois_k5_hetm}
\end{figure}

\begin{figure}[!htbp]
\centering
\begin{subfigure}{0.45\textwidth}
  \centering
  \includegraphics[width=1\linewidth]{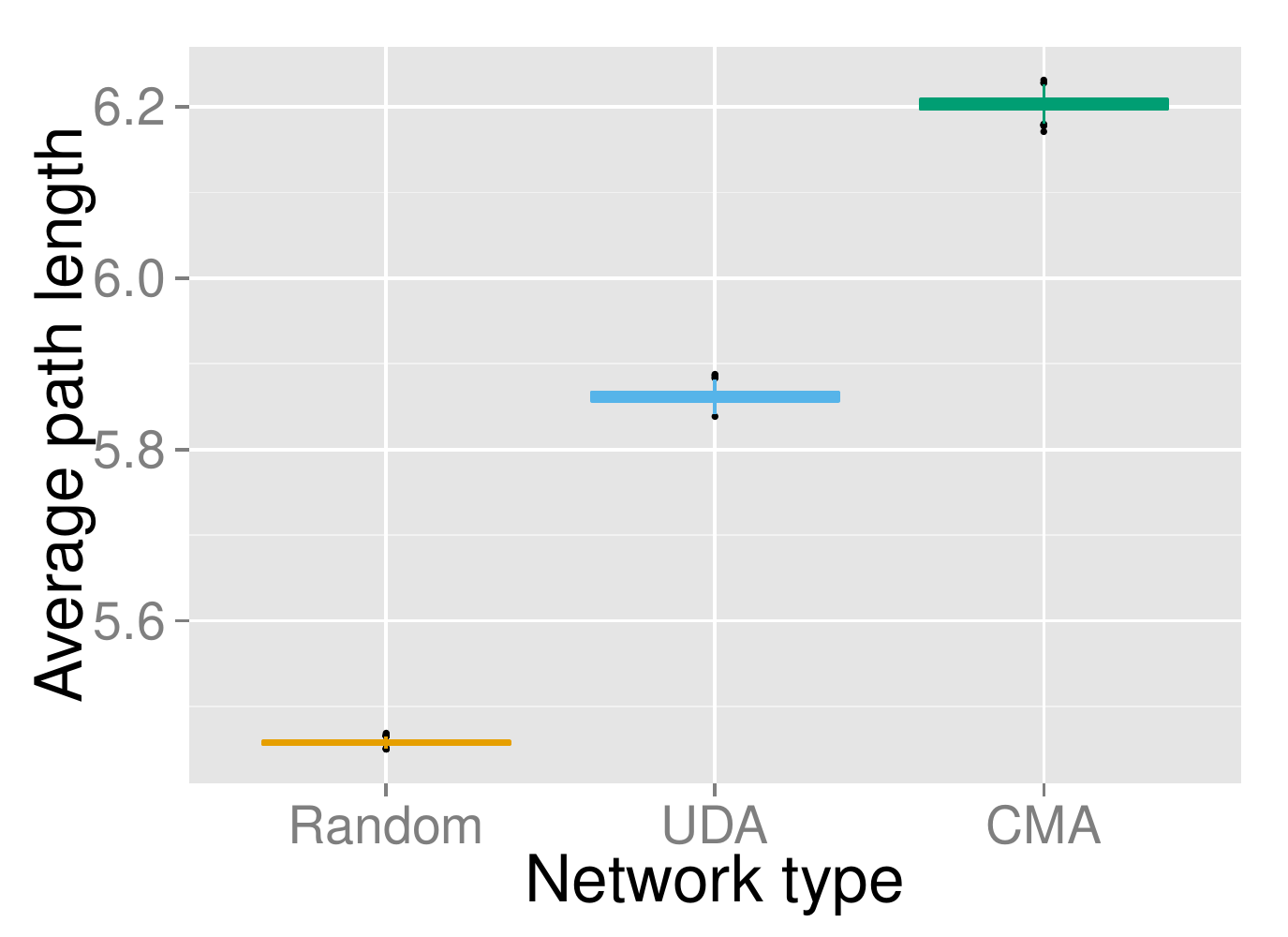}
  \caption{Average path length}
  \label{fig:pois_k5_avpl}
\end{subfigure}%
\begin{subfigure}{0.45\textwidth}
  \centering
  \includegraphics[width=1\linewidth]{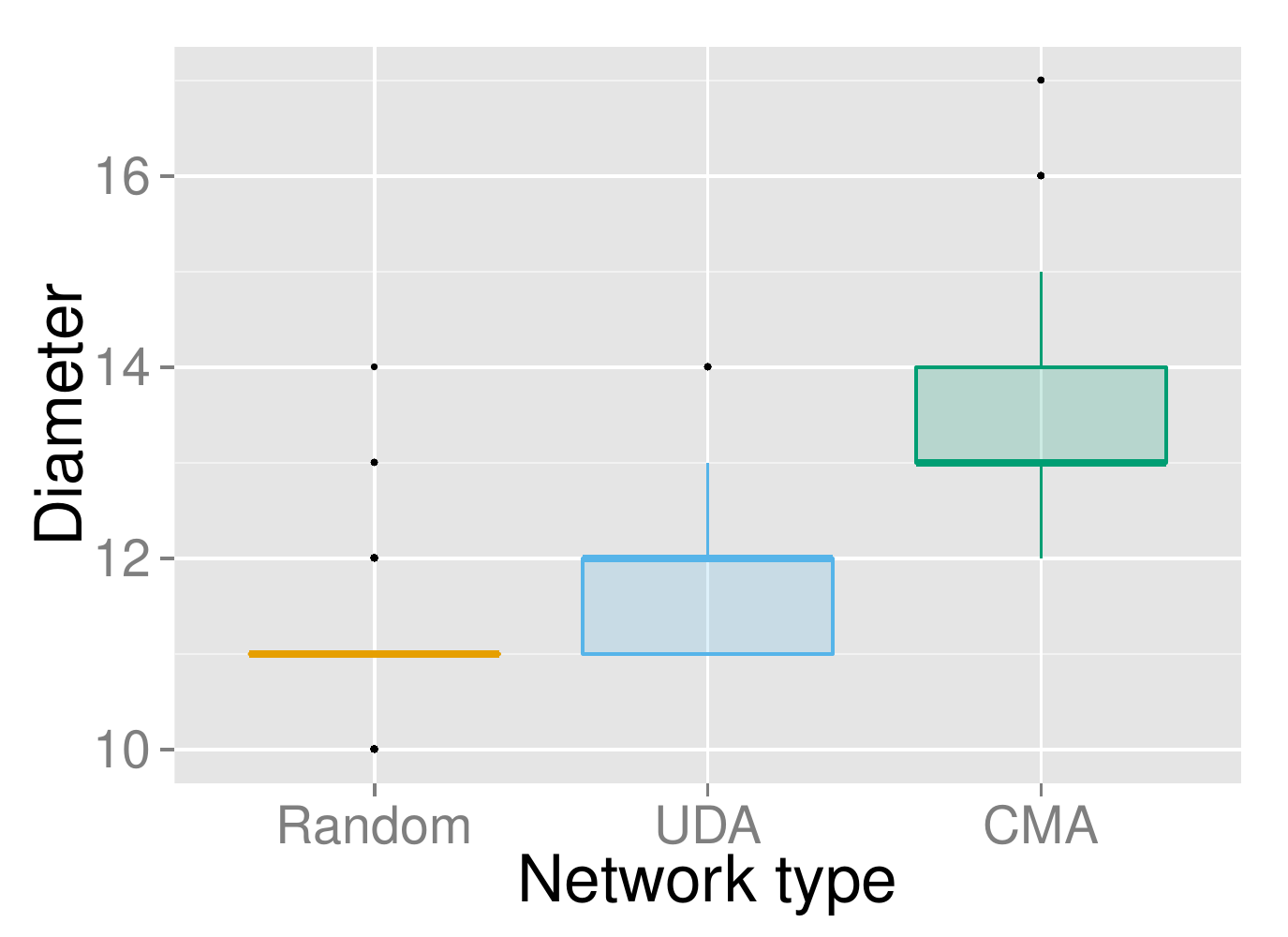}
  \caption{Diameter}
  \label{fig:pois_k5_diam}
\end{subfigure}
\caption{Plots of average path length and diameter for network family \textbf{B} with $k \sim Pois(5)$. The UDA and CMA networks have a global clustering coefficient of $C=0.13$. The increased average path length and diameter between the UDA and random networks is attributable to the higher clustering. The similar increase between UDA and CMA networks is a reflection of the higher assortativity of the CMA networks.}
\label{fig:pois_k5_pl}
\end{figure}

The heterogeneity in degree distribution allows us to use additional degree-dependent metrics: degree-degree correlations and degree-dependent clustering~\cite{newman2002assortative, serrano2005tuning}. These have been plotted in Figure~\ref{fig:pois_k5_hetm}. The plot for the degree-degree correlation coefficient shows that by aggregating clustered subgraphs around high-degree nodes, the CMA-constructed networks yield a higher assortativity than that of UDA and random networks, see Figure~\ref{fig:pois_k5_ass}. This is an important property of the methodology since the clustering potential of a network is bounded by the degree-degree correlation coefficient~\cite{serrano2005tuning}. Moreover, if one wishes to maximise clustering in heterogeneous networks, it is necessary for nodes of similar degree to mix preferentially. Figure~\ref{fig:pois_k5_ddc} shows that the CMA networks yield a negatively skewed distribution of degree-dependent clustering, with nodes of degree $k\geq 9$ contributing most to clustering. The ability to manipulate the degree and clustering relationship as well as assortativity clearly demonstrates the broader scope of the CMA when sampling from the ensemble of networks with same degree distribution and global clustering.   

\begin{figure}[!htbp]
\centering
\begin{subfigure}{0.31\textwidth}
  \centering
  \includegraphics[width=1\linewidth]{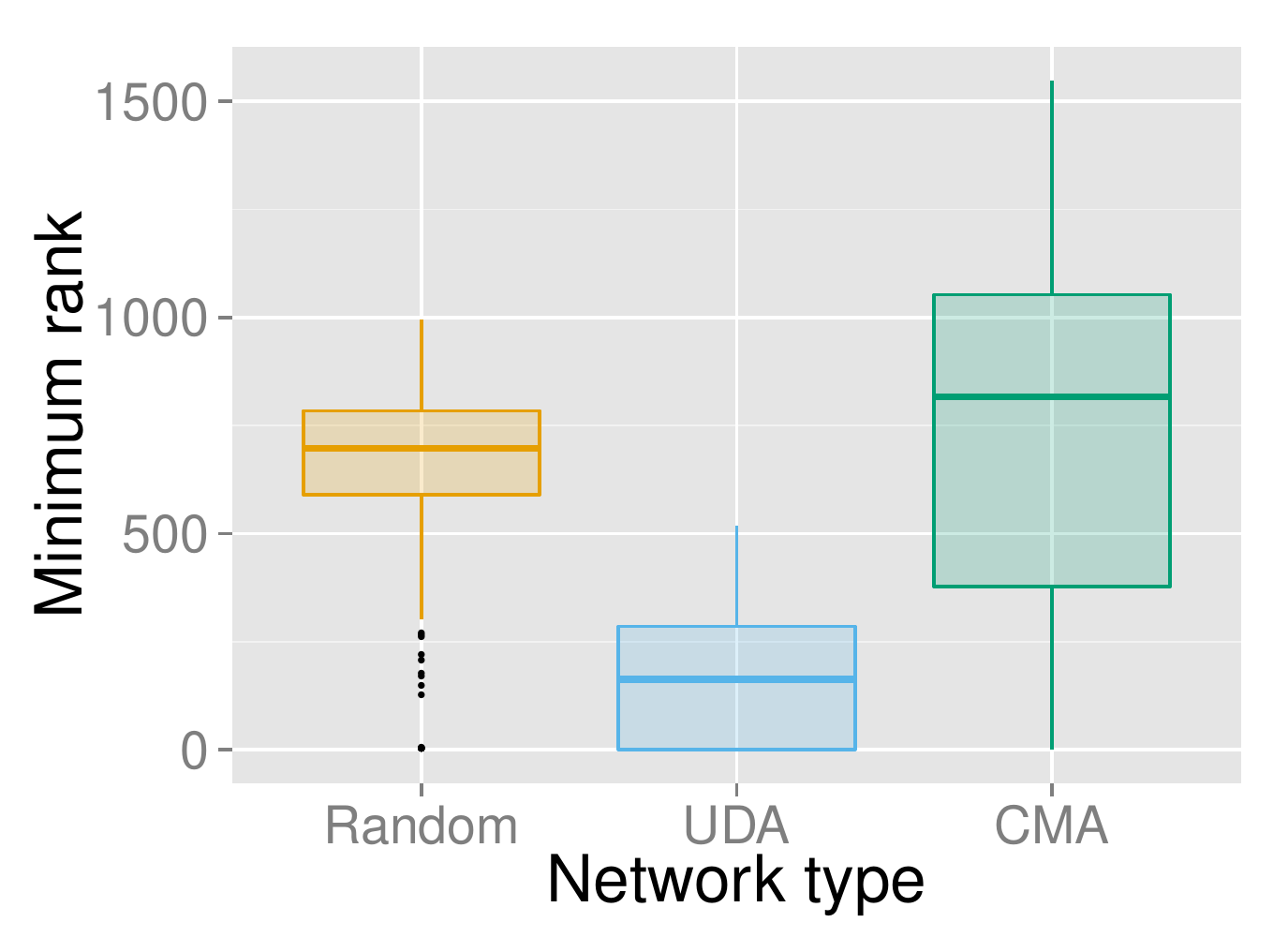}
  \caption{Minimum}
  \label{fig:pois_k5_min_betw}
\end{subfigure}%
\begin{subfigure}{0.31\textwidth}
  \centering
  \includegraphics[width=1\linewidth]{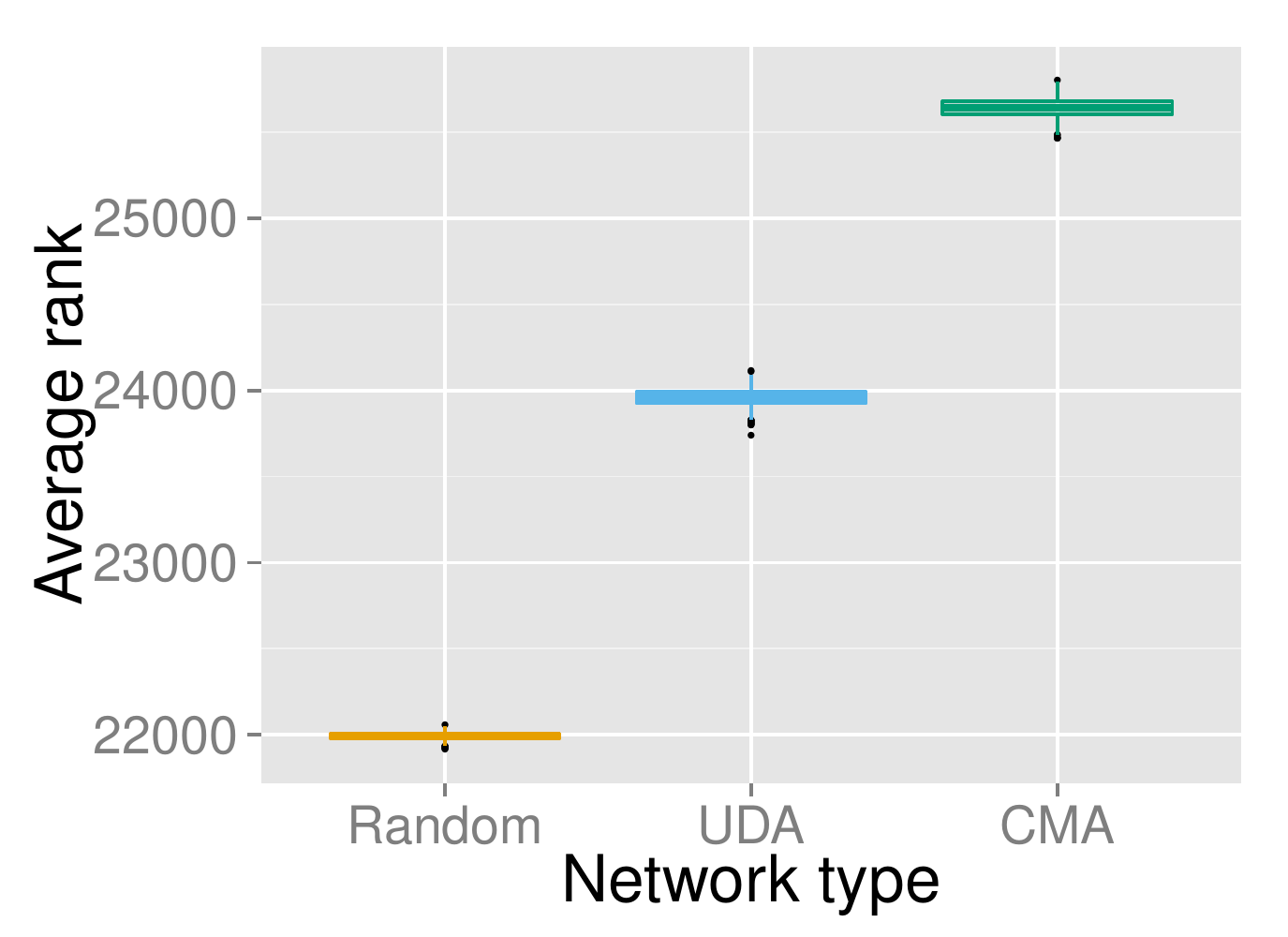}
  \caption{Average}
  \label{fig:pois_k5_avg_betw}
\end{subfigure}
\begin{subfigure}{0.31\textwidth}
  \centering
  \includegraphics[width=1\linewidth]{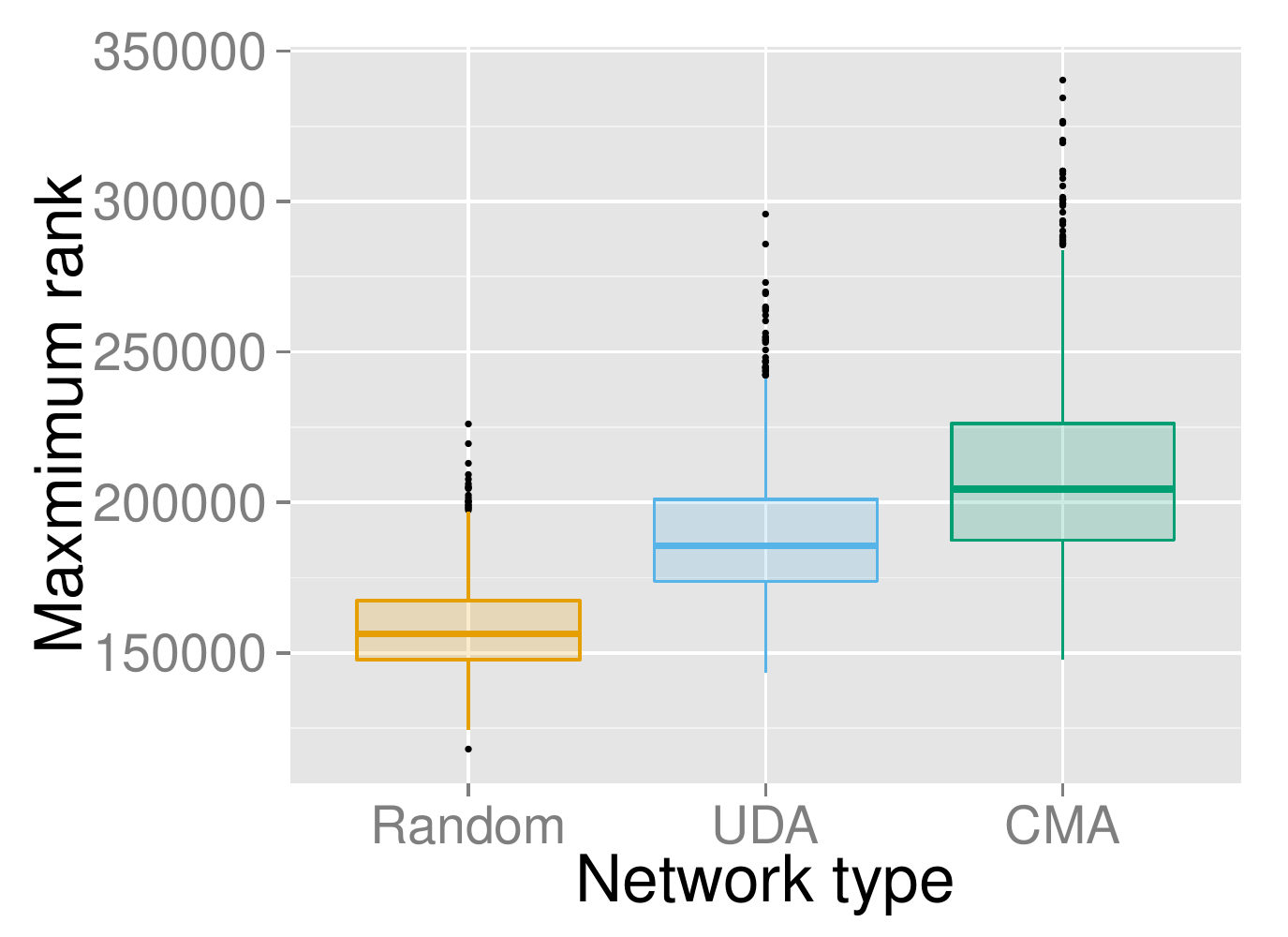}
  \caption{Maximum}
  \label{fig:pois_k5_max_betw}
\end{subfigure}
\caption{Plots of betweenness centrality for network family \textbf{B} with $k \sim Pois(5)$. The UDA and CMA networks have a global clustering coefficient of $C=0.13$. A trend of increasing average and maximum betweenness centrality is observed between random, UDA and CMA networks, respectively.}
\label{fig:pois_k5_betw}
\end{figure}

As with network family \textbf{A}, an increase in average path length, diameter, average and maximum betweenness centrality of UDA and CMA networks over random networks will be attributable to the increased global clustering coefficient, $C=0.13$, see Figure~\ref{fig:pois_k5_pl} and ~\ref{fig:pois_k5_betw}. However, since UDA and CMA networks share the same degree sequence and global clustering coefficient differences in these metrics between UDA and CMA can only be due to increased degree-degree correlation and negatively skewed distribution of degree-dependent clustering. It has previously been noted that increased assortativity corresponds to an increase in average path length \cite{xulvi2004reshuffling} and this will be compounded by the higher-degree nodes (which inevitably serve as central hubs) being more clustered. Similarly, an increase in diameter (a function of path length) will be due to these highly clustered high-degree nodes. Finally, Figures~\ref{fig:pois_k5_avg_betw} and~\ref{fig:pois_k5_max_betw} show a significant increase in average and maximum betweenness centrality between UDA and CMA networks. This is yet another manifestation of the presence of these highly-clustered high-degree nodes.

Table~\ref{tab:countsB} presents a comparison between \emph{measured} and \emph{expected} average subgraph counts for the networks in family \textbf{B}. %Figure~\ref{fig:pois_k5_counts} complements this comparison by displaying the distributions of the four higher-order subgraphs used to construct the networks, namely, $\{G_\triangle, G_\boxempty, G_\boxslash, G_\boxtimes\}$. 
Whereas there is good agreement for UDA networks, it is observed that CMA networks have produced by-products other than what was expected at random, e.g., an additional 50\% $G_\boxslash$ have appeared as by-products. The effects of finite size have been exacerbated by aggregating clustered subgraphs around higher degree nodes, effectively excluding lower to medium degree nodes during this part of the connection process. Within this densely connected component it is easy to envisage a situation where adding only a single edge may create additional (unwanted) subgraphs. This highlights the fact that whilst the total number of $G_\triangle$ is preserved (as evidenced by identical global clustering), the way these subgraphs contribute to higher-order structure can vary significantly.

\begin{table}[!htbp]
\centering
\begin{tabular}{ccccc}\rowcolor{LightBlue} 
   & c4 & d4 & e4 & t3 \\ \hline 
Random & 0   & 0 & 79 & 21 \\ \rowcolor{LightBlue}
UDA & 243   & 504 & 587 & 718 \\
CMA & 232  & 743 & 772 & 691\\ \rowcolor{LightBlue}
Expected & 243  & 504 & 619 & 741
\end{tabular}
\caption{Subgraph counts for network  \textbf{B} ($N=5000$, $k \sim Pois(5)$ and $C=0.13$). The counts are unique. The expected counts are computed by summing the total counts from the subgraph sequences, dividing them by the subgraphs' node cardinality, and adding these figures to the number of subgraphs found as by-products in the random network. The counts for $t3$ are for $G_\triangle$ subgraphs that do not appear in any other subgraphs.}
\label{tab:countsB}
\end{table}

This Section has highlighted that control over the choice of subgraph families and their distributions makes it possible to flexibly explore the solution space of networks with the same degree distribution and global clustering. This in turn provides us with the means to investigate specific areas of this solution space as well as further our understanding of how network metrics deal with such diversity.

\subsection{Does higher-order structure matter?}\label{sec:SIS}

In order to answer this question we make use of the network families \textbf{A} and \textbf{B} detailed above and test the impact of higher-order structure by considering the outcome and evolution of widely used dynamics on networks, namely, $SIS$, $SIR$ and the complex contagion model.  

For each network type in families {\textbf A} and {\textbf B} a series of networks were generated. For each network we performed a single Gillespie realisation of the $SIS$, $SIR$ and complex contagion epidemics. The mean time evolution of infectious prevalence was then calculated, plotted and compared between network types. Complex contagion dynamics was simulated in a similar way but without recovery and remembering that a single infectious contact was usually not sufficient to result in an infected node. Different thresholds of infection and infectious seeds were used and these are specified in figure captions. Matlab code for the $SIS$ and $SIR$ Gillespie algorithms is available from \url{https://github.com/martinritchie/Dynamics}.

%%-------------------------------------------------------------
%% SIS and SIR dynamics
%%-------------------------------------------------------------

\begin{figure}[!htbp]
\centering
\begin{subfigure}{0.45\textwidth}
  \centering
  \includegraphics[width=1\linewidth]{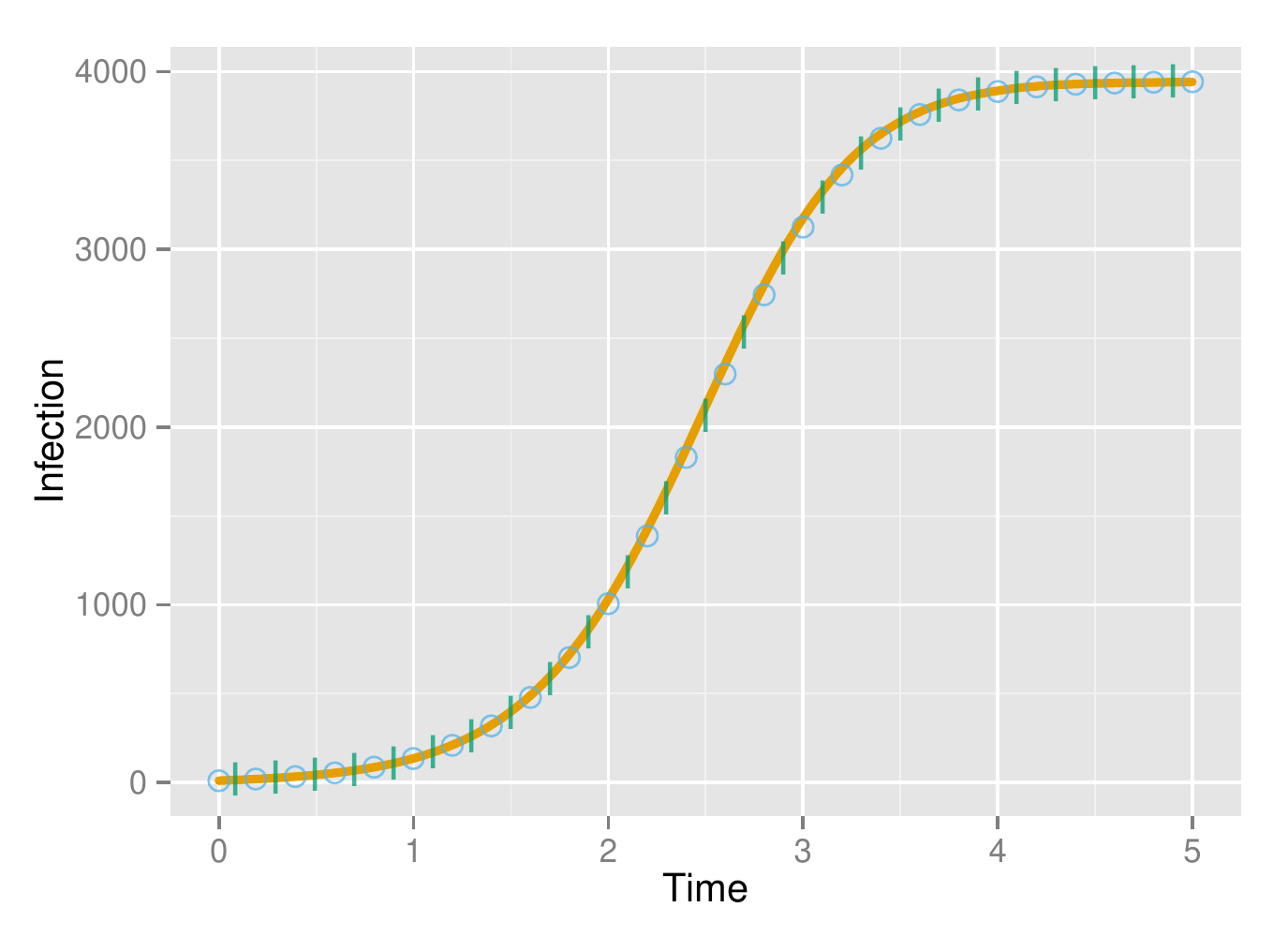}
  \caption{SIS}
  \label{fig:k5_homo_cycles_SIS}
\end{subfigure}%
\begin{subfigure}{0.45\textwidth}
  \centering
  \includegraphics[width=1\linewidth]{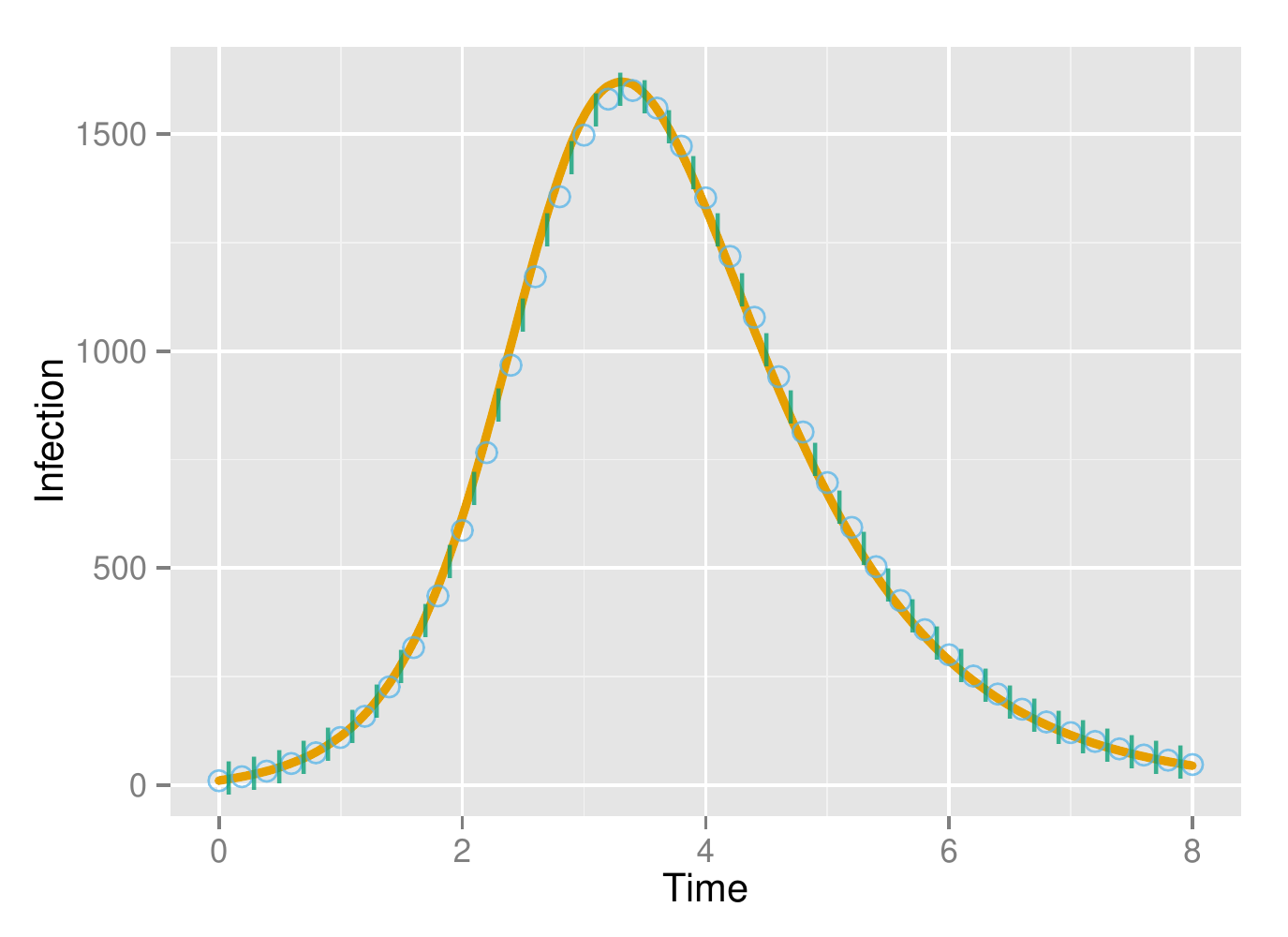}
  \caption{SIR}
  \label{fig:k5_homo_cycles_SIR}
\end{subfigure}
\vskip\baselineskip
\begin{subfigure}{0.9\textwidth}
  \centering
  \includegraphics[width=1\linewidth]{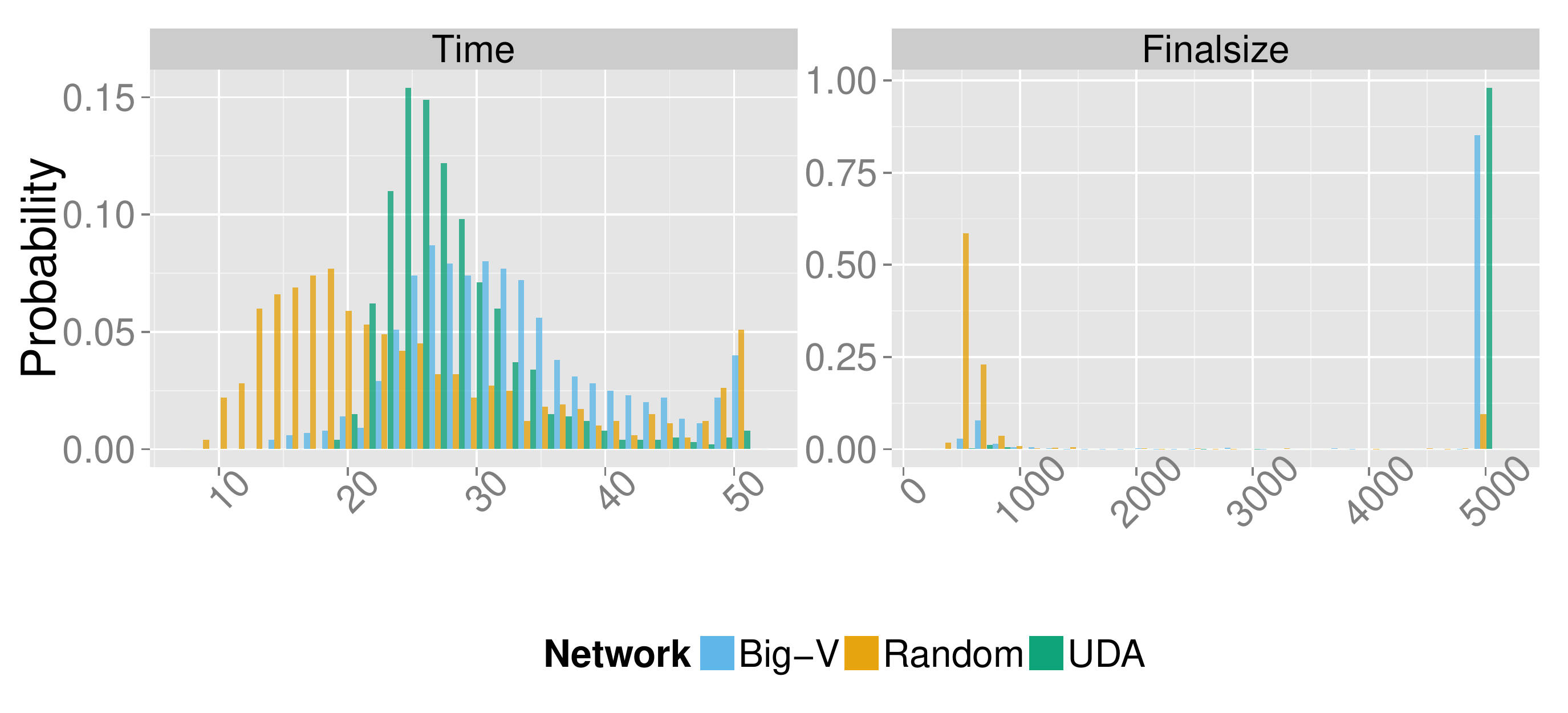}
  \caption{Complex contagion}
  \label{fig:homo_k5_cycles_cc}
\end{subfigure}
\caption{Epidemic dynamics for network family \textbf{A} with $k=5$. The Big-V and CMA networks have a global clustering coefficient of $C=0.04$. In (a) and (b) the orange line, blue circles and green vertical markers correspond to the random, Big-V and UDA networks respectively. In (c) and (d) the same colour scheme is used but with bars. The $SIS$ and $SIR$ epidemics represent the average of single Gillespie simulations on each of the 1000 network realisations from each network generation algorithm. The $SIS$ and $SIR$ epidemics were seeded with an initial infectious seed of $I_0=10$ and had a per link rate of infection of $\tau=1$ and recovered independently at rate $\gamma=1$. The complex contagion epidemics had an initial infectious seed of $I_0=250$ and a fixed threshold of infection of $r=2$.}
\label{fig:k5_homo_cycles_epi}
\end{figure}

\begin{figure}[!htbp]
\centering
\begin{subfigure}{0.45\textwidth}
  \centering
  \includegraphics[width=1\linewidth]{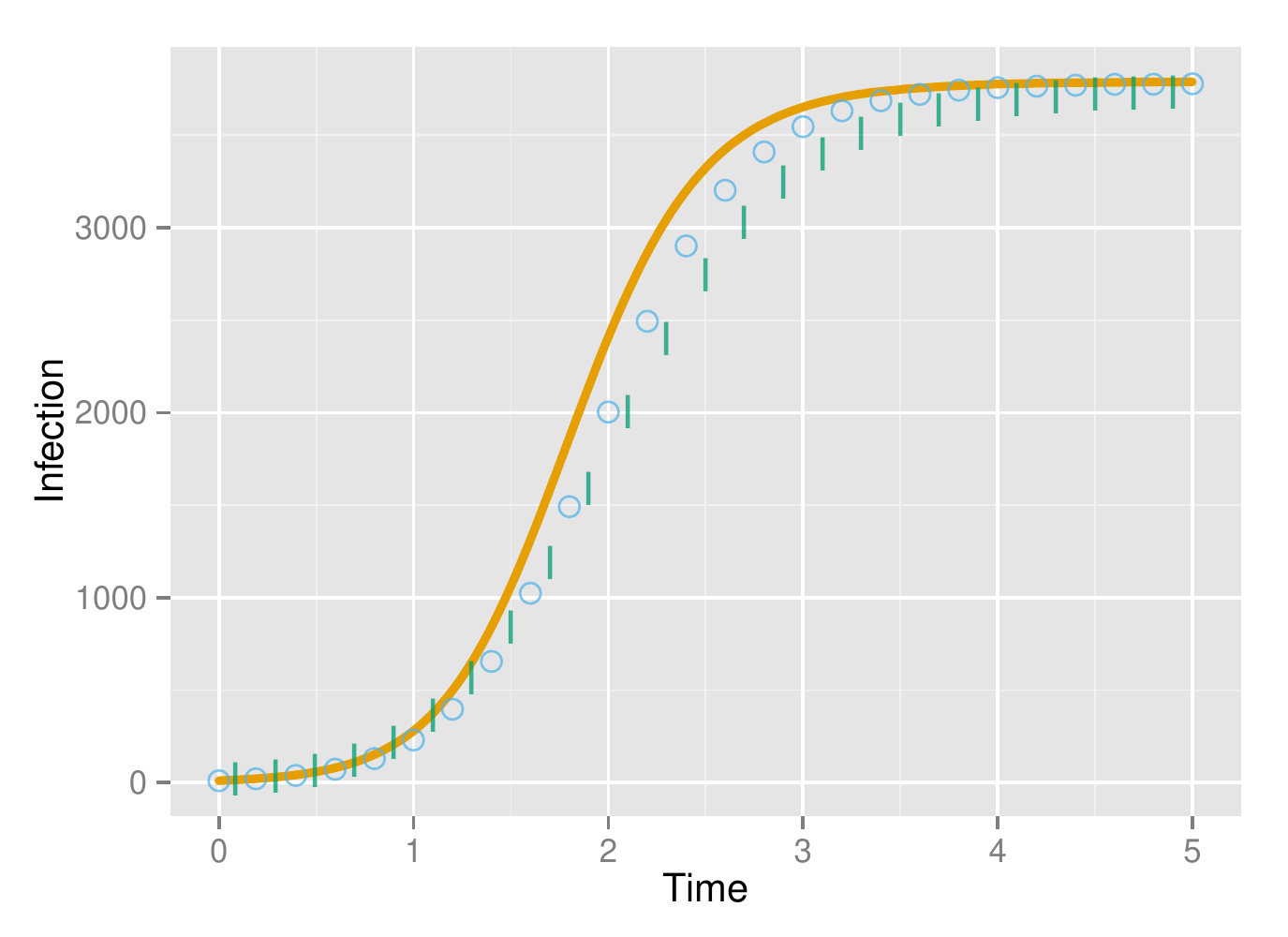}
  \caption{SIS}
  \label{fig:pois_k5_sis}
\end{subfigure}%
\begin{subfigure}{0.45\textwidth}
  \centering
  \includegraphics[width=1\linewidth]{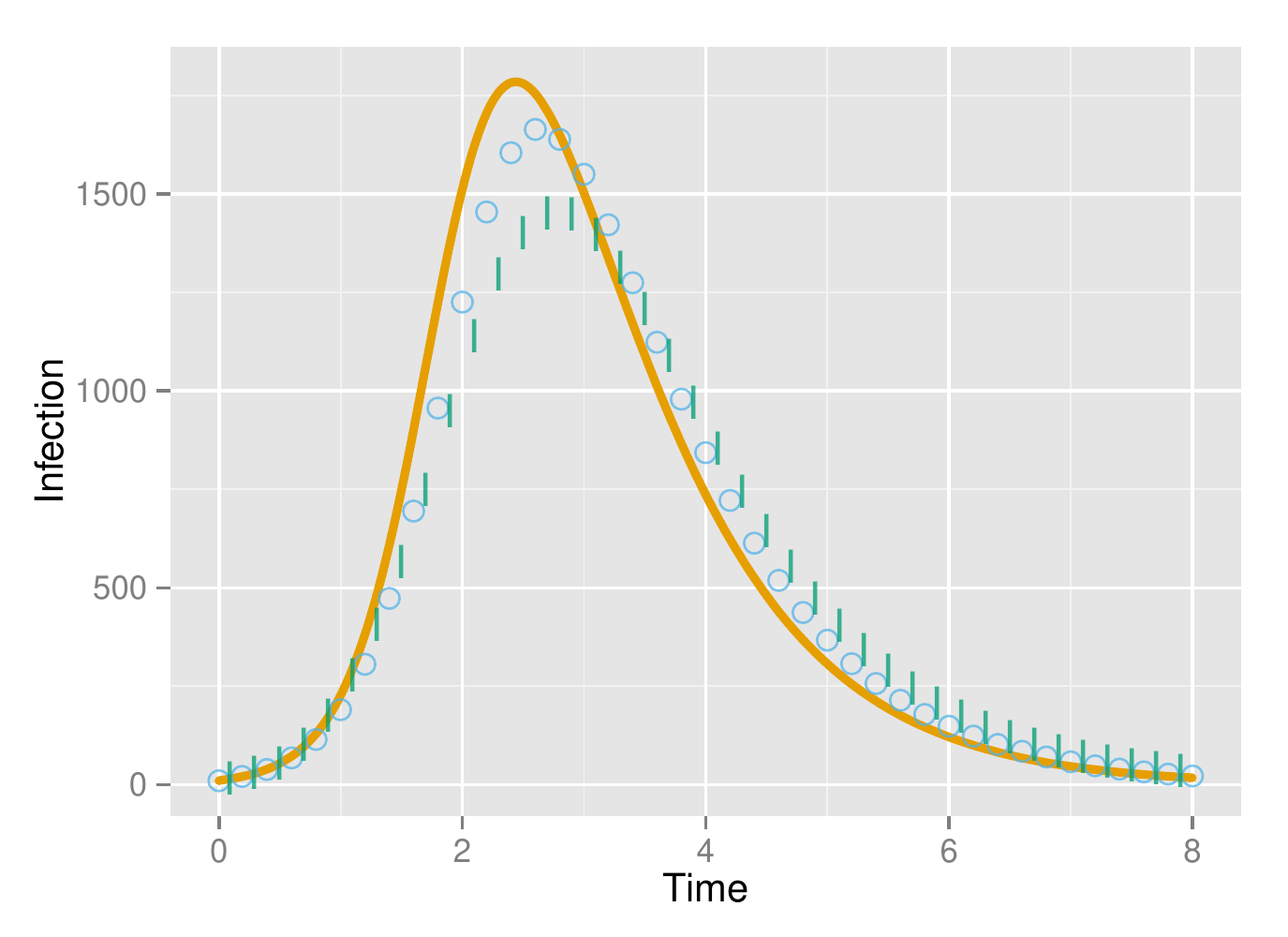}
  \caption{SIR}
  \label{fig:pois_k5_sir}
\end{subfigure}
\vskip\baselineskip
\begin{subfigure}{0.9\textwidth}
  \centering
  \includegraphics[width=1\linewidth]{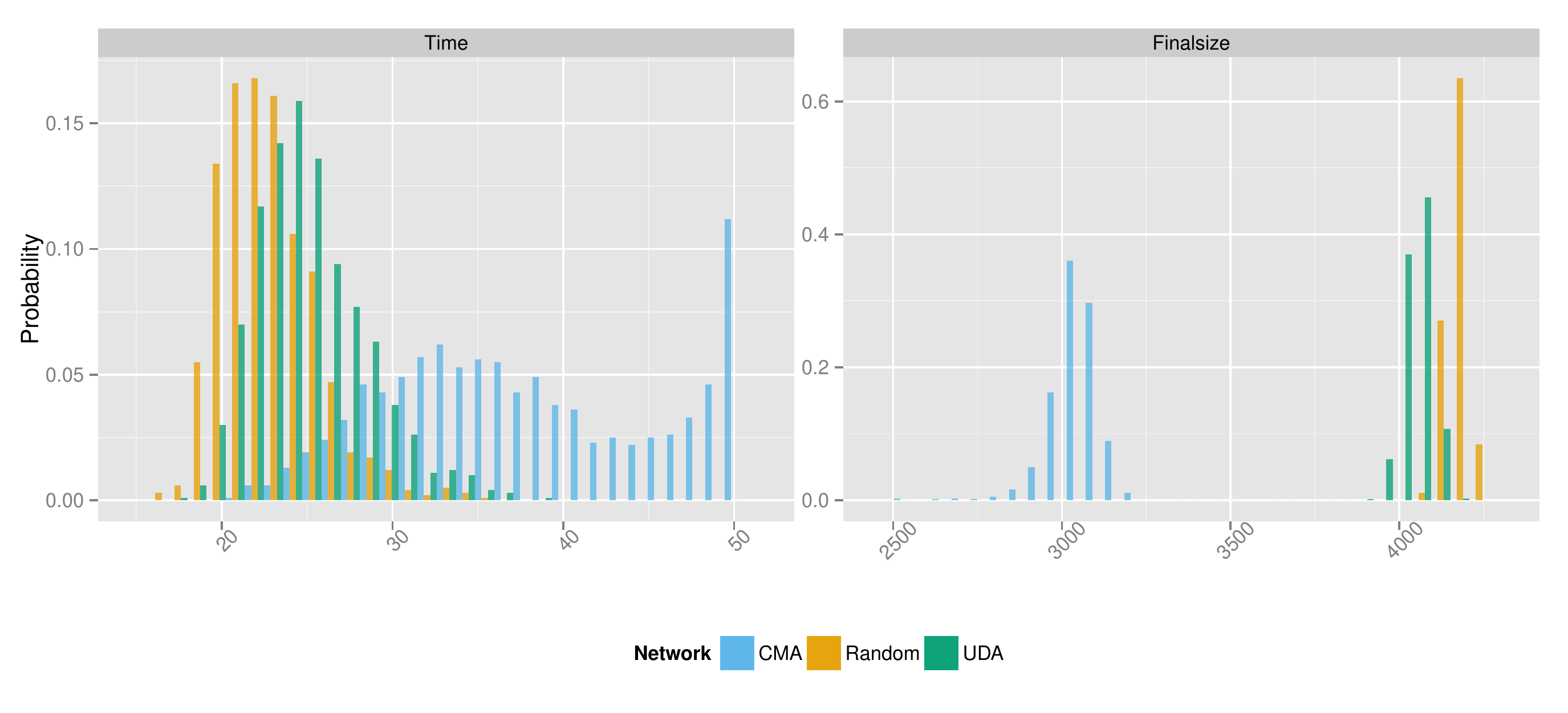}
  \caption{Complex contagion}
  \label{fig:pois_k5_classical_cc}
\end{subfigure}
\caption{Epidemic dynamics for network family \textbf{B} with $k \sim Pois(5)$. The UDA and CMA networks have a global clustering coefficient of $C=0.13$. In (a) and (b) the orange line, blue circles and green vertical markers correspond to the random, UDA and CMA networks respectively. In (c) and (d) the same colour scheme is used for the bars. The $SIS$ and $SIR$ epidemics represent the average of single Gillespie simulations on each of the 1000 network realisations from each network generation algorithm. The $SIS$ and $SIR$ epidemics were seeded with an initial infectious seed of $I_0=10$ and had a per link rate of infection of $\tau=1$ and recovered independently at rate $\gamma=1$. The complex contagion epidemics had an initial infectious seed of $I_0=1000$ and a fixed threshold of infection of $r=3$. In the $SIS$ and $SIR$ dynamics the effect of clustering on the high degree nodes, inhibiting disease propagation, dominates the epidemiological encouraging effect of increased assortativity.}
\label{fig:pois_k5_epi}
\end{figure}

We know by construction that members of network family \textbf{A} were generated using different subgraphs and Section~\ref{sec:UDA_CMA} has shown that observable differences were found between networks in terms of average path length, betweenness centrality and subgraph composition. Despite this, Figures~\ref{fig:k5_homo_cycles_SIS} and~\ref{fig:k5_homo_cycles_SIR}, which show the time evolution for $SIS$ and $SIR$ dynamics respectively, illustrate that these dynamics can display a certain degree of insensitivity to these differences in structure. In this case, it is the $SIR$ dynamics that show the greatest difference, in peak infectious prevalence (Figure~\ref{fig:k5_homo_cycles_SIR}) albeit quite marginal. In contrast, complex contagion dynamics do show sensitivity to structural differences found between Big-V and UDA networks. Figure~\ref{fig:homo_k5_cycles_cc} reveals that for Big-V networks the epidemic fully percolates in almost 100\% of the simulations instead of only 80\% of the cases for UDA networks. This indicates that whilst Big-V networks operate in the super critical regime, UDA networks are closer to the transition point. Locating this transition is possible but is beyond the scope of this paper. 

When network family \textbf{A} is used, the networks' degree distribution and clustering appear to be the main determinants of the time evolution and outcome of the $SIS$ and $SIR$ epidemics. In contrast, when network family \textbf{B} is used, Figure~\ref{fig:pois_k5_epi} shows that all dynamics considered are impacted by differences in network topology. For Figures~\ref{fig:pois_k5_sis} and~\ref{fig:pois_k5_sir}, a trend of inhibited spread of infection is observed from the random to UDA to CMA networks. It has already been shown that clustering slows the spread of infection~ \cite{keeling1999effects,green2010large}, and we see that this effect dominates over higher assortativity, which usually leads to faster initial spread of the epidemic \cite{kiss2008effect}. Similarly, Figure~\ref{fig:pois_k5_classical_cc} which shows the distribution of the final epidemic size for the complex contagion dynamics reveals that: (a) the higher clustering observed in the UDA networks fails to have a significant impact when compared to the random network equivalent and (b) the CMA networks significantly slow the pace of the epidemic as well as reduce its final size compared to both random and UDA networks. Hence, for the UDA and CMA networks where both degree distribution and global clustering are identical the observed differences are explained by the combined effect of varying distributions of subgraph around nodes and varying prevalence of subgraphs (both of which are related to one another to some extent) as shown by %Figure~\ref{fig:pois_k5_counts} and 
    Table~\ref{tab:countsB}.

Taken together, our simulation data shows that even though the proposed algorithms construct networks with identical degree sequence and global clustering, these networks can give rise to measurable differences in resulting epidemics, be it it time evolution or final outcome. With the exception of $SIS$ and $SIR$ epidemics on network family \textbf{A} (still with some small differences) we found significant differences in all other instances. A more systematic investigation of more network models and wider parameter range for the dynamics is needed but left to future work.

\section{Discussion}\label{sec:discussion}

In this paper, we have described two novel network generating algorithms that strictly preserve a given degree sequence whilst permitting control over the building blocks of the network and enabling tuning of global clustering. We have compared these algorithms to one another as well as to the widely used Big-V rewiring algorithm. Using our algorithms we have empirically demonstrated that it is possible to create networks that are identical with respect to degree sequence and global clustering, yet elicit significant differences in network metrics and in the outcome of dynamical processes unfolding on them. We have presented evidence to suggest that the methods sample from different areas of the network state space and that these sampling variations do matter. 

Of the two algorithms proposed, UDA is the simplest to use. It requires less input and is conceptually elegant. We believe that this algorithm, when parametrised with complete subgraphs, would more likely yield analytical results due to its combinatorial nature. Note that whilst varying levels of clustering can be achieved and estimated before network construction it is not possible to target a specific level of clustering, due to the emergent nature of the distribution of subgraphs around nodes. A second potential limitation of this algorithm is its dependence on solving the underdetermined Diophantine equations that reside in high dimensional spaces. Computationally, it may become difficult to include large families of subgraphs. However, we did not encounter such problem in our experiments.

The CMA algorithm is more complex but also more versatile. Being able to specify distributions of subgraphs alongside a given degree sequence, and preserve both, is highly novel. Fixing the degree sequence allows for some interesting ways to construct the subgraph sequences. Knowing the number of nodes in each degree class $k$ allows us to combine such nodes to form complete subgraph with $k$ nodes, as this requires ($k-1$) links. The remaining single edges can then be used to connect to the rest of the network. We used this for the heterogeneous degree sequence presented in the results section, yielding a network with a global clustering coefficient of $C=0.67$ and a giant component of $N\approx 4800$ out of $N=5000$ nodes. We were unable to achieve such high clustering with either Big-V or UDA algorithms. It must be noted that in our application of this algorithm, we had the luxury of using hyperstubs sequences we knew to be graphical -- them being output from the UDA -- to guide how we parametrised the CMA. In general, the stub sequences induced by the hyperstub sequences would have to be constrained to ensure that they are graphical. This is possible but must be taken into account when considering applying this algorithm. 

We have shown that despite identical degree distribution and global clustering, significant diversity in networks can still be elicited. This has occurred in two ways: (1) by construction, by redistributing the same number of subgraphs and (2) unexpectedly, through the emergence of by-products. We conjecture that any controlled -- or believed to be controlled -- network generation algorithm will yield by-products, unless heuristic constraints are introduced to reduce the likelihood of subgraphs sharing lower-order subgraph components for example. As witnessed in our results, even configuration model networks lead to a large number of loops with 4, 5, and 6 nodes (longer cycles were not measured). This problem can only be exacerbated when control of more sophisticated structures is implemented. As such, care has to be taken when parametrising algorithms. For example, one would need to specify a relatively large number $G_{\hexagon}$ subgraphs in a network's construction to impact the subgraph count beyond what one would observe by chance in a random network. More surprisingly, as we witnessed with $G_\boxslash$ subgraphs in the CMA networks from network family \textbf{B}, significant numbers of subgraph by-products can appear in addition to what was observed in the random networks depending on how one wishes to place the subgraphs around nodes.  

We have seen that by using a very modest selection of subgraphs, we have been able to substantially influence dynamics running on the network, particularly complex contagion dynamics. All results relating to this model indicate that constraining a network by degree sequence and clustering is not sufficient to accurately predict the course of the epidemic. More importantly, the results appear to suggest that the location of the critical regime depends on the higher-order structure of the network (above and beyond clustering). 

Being able to generate networks with different structural properties or higher-order structure is a key feature of any network construction algorithm. However, if such structural details do not impact on dynamics unfolding on the network, then models for such dynamics can rely with high confidence on a limited set of network descriptors. Although degree sequence, degree-degree correlations and global clustering coefficient were observed to be the main drivers of disease transmission in models such as $SIS$ and $SIR$, we found it not to be true in general. This is an important finding because one should remember that the dynamics simulated here are modest in complexity, when compared to models of neuronal dynamics for example, and yet, we were able to elicit significant differences by simply tuning the network structure above and beyond triangles. This implies that determining the role and impact of higher-order structure may yet hold many important and surprising answers.  \\

\noindent{\textbf{Acknowledgements:}} Martin Ritchie gratefully acknowledges EPSRC (Engineering and Physical Sciences Research Council) and the University of Sussex for funding for his PhD. We would also like to thank Dr J.C. Miller for useful discussions on the complex contagion model~\cite{miller2015complex}, and for sharing his code for simulating the complex contagion model on networks~\cite{miller2015private}.

%\begin{figure}[!htbp]
%\centering
%\includegraphics[scale=1]{pois_k5_class_sgcount_new}
%\caption{Distributions of total number of subgraphs in network family \textbf{B} ($N=5000$, $k \sim Pois(5)$). The UDA and CMA networks have a global clustering coefficient of $C=0.13$. All given counts are unique. The $t3$ counts denote the number of $G_\triangle$ subgraphs that are not involved in any subgraphs of four nodes (i.e., $G_\boxslash$ and $G_\boxtimes$). However, the $c4$ and $d4$ counts may include $G_\triangle$ subgraphs shared by $G_\boxslash$ and $G_\boxtimes$.}
%\label{fig:pois_k5_counts}
%\end{figure}

\clearpage

\section{Appendix}\label{sec:app}

\subsection{Pseudocode for UDA}\label{UDA}
\begin{algorithm}[!htbp]
\SetKwInOut{Input}{input}\SetKwInOut{Output}{output}
%\SetKwProg{myfun}{Function}{}{}
%\HiLio\myfun{UDA}{
\HiLiy\Input{$D=(d_1,d_2,\dots,d_N)$, $G=\{G_1,G_2,\dots,G_l\}$}
\HiLiy\Output{$H \in \mathbb{N}_0^{l \times N}$.}
\HiLib\KwSty{Variables} \\
$D$: degree sequence, $N$: number of nodes, \\
$G$: set of subgraphs, $l$: number of subgraphs, \\
$g_i$: subgraph adjacency matrix, $X_k$: solution space for degree $k$, \\
$H$: hyperstub degree sequence \\
\HiLio\KwSty{Procedure} \\
\For{Each subgraph, $G_i$}{
\HiLig{\% Identify the degree sequences of the subgraphs.} \\
$s_i = \sum g_i$ \\
\HiLig{\% Take the unique elements.} \\
$s_i = unique(s_i)$} 
\HiLig{\% Concatenate into a single vector.} \\
 $S = (s_1, s_2, \dots s_l)$ \\ 
 \For{$k=1,2,\dots k_{max}$}{
 \HiLig{\% $X_k(i,:)$ denotes a hyperstub arrangement for a degree $k$ node.} \\
 $X_k = diorecur(S,k)$ \\
 }
 \For{n = 1,2,\dots, N}{
 \HiLig{\% Take random element from the solution space.} \\
 $r = rand$; $h_n =  X_{D(n)}(r,\cdot)$}
 \HiLig{\% Concatenate into a single matrix.} \\
 $H = (h_1, h_2, \dots, h_l)$ \\ 
 \nl \KwRet
%}
\caption{Pseudocode for the underdetermined network generation algorithm (UDA). This pseudocode focuses on the salient points of the UDA, namely, how the algorithm draws solutions from the solution space of an underdetermined Diophantine equation to determine the arrangement of hyperstubs around a particular node. Other steps, such as ensuring the handshake lemma is satisfied for both lines and subgraphs, are detailed in Section~\ref{sec:UDA} and can be viewed in the source code. The output hyperstub degree sequence $H$ must be used as input for a modified configuration model connection process to realise a network, see Section~\ref{sec:CP}.}\label{algo:UDA}
\end{algorithm}

\clearpage

\subsection{Pseudocode for CMA}\label{CMA}
\IncMargin{1em}
\begin{algorithm}[!htbp]
\SetKwInOut{Input}{input}\SetKwInOut{Output}{output}
%\SetKwProg{myfun}{Function}{}{}
\SetKwFunction{FRecurs}{FnRecursive}
\HiLiy\Input{$D=(d_1,d_2,\dots,d_N)$, $G=\{G_1,G_2,\dots,G_l\}$, $S=\{S_1,S_2,\dots,S_l\}$.}
\HiLiy\Output{$H \in \mathbb{N}_0^{|s| \times N}$.}
\HiLib\KwSty{Variables} \\
$D$: degree sequence, $N$: number of nodes, \\
$G$: set of subgraphs, $l$: number of subgraphs, \\
$S$: subgraph sequence, $g_i$: subgraph adjacency matrix, \\
$|s|$: number of unique corners in a subgraph, $H$: hyperstub degree sequence \\
%\HiLio\myfun{CMA}{
\HiLio\KwSty{Procedure} \\
\For{Each subgraph, $G_i$}{
\HiLig{\% Identify the degree sequence, $s$, of the subgraph.} \\
$s_i = \sum g_i$, $s_i = unique(s_i)$, $m = length(s_i)$ \\
\HiLig{\% $p$ reflects the proportions of hyperstubs} \\
$p_i = (p_1,p_2,\dots,p_m)$ \\
	\For{$j = 1,2,\dots,N$}{
    \HiLig{\% The subgraph sequence is decomposed into a hyperstub} \\
    \HiLig{\% sequence using the multinomial distribution, $M$, } \\
    \HiLig{\% so that $H_i \in \mathbb{N}_0^{m \times N}$  } \\
    $H_i(j) = M(S_i(j),p_i)$, \\}
    \HiLig{\% $H'_i$ is a sequence of the true stub count } \\
    $H'_i=H_i \cdot s_i$\\
    \HiLig{\% Sum so that $H'_i \in \mathbb{N}_0^{1 \times N}$ } \\
    $H'_i(j) = \sum_{\alpha=1}^{m}H'_i(\alpha,j)$ \\
}
\While{elements of each $H_i$ are non-zero}{
\HiLig{\% Find the largest subgraph degree, } \\
$h_i(j) = \max \{\max \{H'_1\}, \max \{H'_2 \}, \cdots, \max \{H'_l\} \}$ \\
\HiLig{\% i.e., the $j^{th}$ element of $H_i$. } \\
\HiLig{\% Find all elements of the degree sequence at least this large and } \\
\HiLig{\% select an element from $d'$ at random} \\
$d' = \{d \in D: d \geq m \}$, $\delta = d'(random)$ \\
\HiLig{\% pair $H_i(j)$ to $\delta$ and update} \\
\HiLig{\% $\delta$'s available degree and $H_i$} \\
$\delta = \delta - H_i(j)$,~ $H_i(j) = 0$ \\
}
%}
\caption{Pseudocode for the cardinality matching algorithm (CMA). Other steps, such as ensuring the handshake lemma is satisfied for both lines and subgraphs, are identical to what is used for the UDA and are detailed in Section~\ref{sec:UDA} and can be viewed in the Matlab source code. The output hyperstub degree sequence $H$ must be used as input for a modified configuration model connection process to realise a network, see Section~\ref{sec:CP}.}
\end{algorithm}

\clearpage

\bibliography{M_Ritchie}{}
\bibliographystyle{unsrt}
\end{document}